\documentstyle[12pt]{article}

\begin{document}
\begin{flushright}
{PITT-94-07}\\
{LPTHE-94-31}\\
{CMU-HEP-94-23}\\
{DOR-ER/40682-77}\\
{7-23-94}
\end{flushright}
\begin{center}
{\bf  DISSIPATION VIA PARTICLE  PRODUCTION}
{\bf IN SCALAR FIELD THEORIES}
\end{center}
\begin{center}
{\bf D. Boyanovsky$^{(a)}$, H.J. de Vega$^{(b)}$ , R. Holman$^{(c)}$,}\\
{\bf D.-S. Lee$^{(a)}$ and  A. Singh$^{(c)}$}
\end{center}
\begin{center}
{\it (a)  Department of Physics and Astronomy, University of
Pittsburgh, Pittsburgh, PA. 15260, U.S.A.} \\
{\it (b)  Laboratoire de Physique Th\'eorique et Hautes Energies$^{[*]}$
Universit\'e Pierre et Marie Curie (Paris VI),
Tour 16, 1er. \'etage, 4, Place Jussieu
75252 Paris, Cedex 05, France}\\
{\it (c) Department of Physics, Carnegie Mellon University, Pittsburgh,
PA. 15213, U. S. A.}
\end{center}
\begin{abstract}
The non-equilibrium dynamics of the first stage of the reheating process,
 that is dissipation via
particle production is studied in  scalar field theories in the unbroken and
in the broken symmetry phase.  We begin with a perturbative study to one loop
and
show explicitly that the mechanism of dissipation via particle production
cannot
be explained with a simple derivative term in the equation of motion. The
dissipative contribution is non-local and there does not exist a local
(Markovian)
limit at zero temperature.  Furthermore, we show that  both an amplitude as
well as a one-loop
calculation present instabilities, requiring a non-perturbative resummation.
Within the same approximations, we study an O(2) linear sigma model that
allows to study  dissipation by Goldstone bosons. We find infrared divergences
that require non-perturbative resummation in order to understand the long-time
dynamics.

We obtain a perturbative Langevin equation that exhibits a generalized
fluctuation-dissipation relation, with non-Markovian kernels and colored noise.

We then study a Hartree approximation and clearly  exhibit dissipative effects
related to the thresholds to particle production. The asymptotic dynamics
depends
on the coupling and initial conditions but does not seem to lead to exponential
relaxation.

The effect of dissipation by Goldstone bosons is
studied non-perturbatively in the
large N limit in an O(N) theory. Dissipation produced by
 Goldstone bosons  dramatically changes the picture of the
phase transition. We find the remarkable result that for ``slow-roll'' initial
conditions (with the expectation value of the field initially near the origin)
the final value of the expectation value of the scalar field is
very close to its initial value. Most of the potential energy is transferred to
the
production of long-wavelength Goldstone bosons.  We find that the minima of
the effective action (as determined by the final value of the expectation value
of
the scalar field) depend on the initial conditions. This in fact points out
that in the case of broken symmetry, in the large N limit there are {\it many}
extrema of the  effective action.
In the Hartree and large N approximation dissipation occurs in a collisionless
regime, a situation very similar to that of Landau damping.
We provide extensive numerical analysis of the
dynamics.
\end{abstract}

\section{Introduction}

It is well appreciated that the dynamics of dissipation in scalar field
theories is of great importance in a variety of settings. One of the most
interesting of these is that of reheating of the universe after an inflationary
epoch has passed\cite{inflation,reviews}. Recall that at the end of new or
chaotic
inflation\cite{newchaotic}, defined by when the slow-roll conditions for the
so-called inflaton field\cite{kolbturner,lindebook}, $\phi(t)$, fail to obtain,
the
inflaton begins to oscillate about its true ground state.

 Since the inflaton is coupled to lighter fields (fermions or other
scalars) as the scalar field oscillates around the true minimum it decays into
these other particles, and eventually these particles thermalize via collisions
and relax to an equilibrium state at high temperature. It is usually stated
that
such coupling
gives rise to a term of the form $\Gamma \dot{\phi}(t)$, where
$\Gamma$ is the decay rate of the inflaton into the lighter
fields\cite{albrechtetal,abbottwise}.
This term,
which is usually put in by hand as a phenomenological result of the existence
of open decay channels, acts as a friction term in the equation of motion for
the inflaton, and converts the inflaton energy density into that of the lighter
particles during its oscillations. These decay products are then supposed to
thermalize, completing the reheating process\cite{albrechtetal,abbottwise}.

This picture bears closer scrutiny. Two questions that must be  addressed
are: (i) how does a time reversible theory generate dissipative dynamics, and
(ii) can the phenomenological term $\Gamma \dot{\phi}(t)$ be derived
from first principles?

The answer to the first question is relatively well known. To generate
dissipative dynamics from a time reversible theory requires that some
``coarse-graining'' of the field degrees of freedom be done. Essentially, one
must trace out (in the functional integral) degrees of freedom other than the
one that has been deemed important to the dynamics. The tracing operation
turns a closed system (that of {\em all} the field modes) into an open one
(that of the field modes of interest) where the traced out modes now become
the ``environment'' which couples to the remaining modes. Another important
necessary ingredient are non-equilibrium initial conditions.

As envisaged in the original scenarios, the process of reheating consists of
two
different stages. During the first one, potential and kinetic energy of the
scalar field is dissipated by the process of particle production. The second
stage involves the collisional thermalization of these particles reaching a
final equilibrium state at some temperature.

 What we want to  understand in this article is the dynamics of the
 first stage
 in an {\em a priori} fashion, i.e. how to incorporate the effect
of quantum  fluctuations into the dynamics of the time
dependent order parameter, and the process of particle production.

This is the first necessary step to fully understand the reheating process in
its full complexity.

What we do here is the following. Starting from a renormalizable self
interacting scalar field theory  we isolate the expectation value (which by
translational invariance
 only depends on time) and generate its equation of motion taking quantum
fluctuations into account. Since we are motivated by
reheating after an inflationary epoch, we treat the dynamics at zero
temperature
since the temperature is supposed to have red-shifted all the way to zero
during
the inflationary stage.

What makes our approach
different from others that have been proposed (see below) is that we are not
considering an ``in-out'' expectation value, such as would be generated by the
usual method of constructing the effective action. Rather, we use a method that
generates the equations for an ``in-in'' expectation value of the field
operator. This ensures that our equations of motion for the field expectation
value are both
real and causal. This has been done by others, most notably by  Calzetta
and Hu\cite{calzettahu} and Paz\cite{paz}.

After obtaining the renormalized, causal and real equations of motion that
determine the dynamics we provide an extensive numerical analysis of the
equations.

The results from our analysis are quite interesting and we summarize them here.
The equations of motion are  rather complicated integro-differential equations
connecting the field expectation value to the fluctuations of the non-zero
momentum modes. We
begin by obtaining the equations of motion in an amplitude expansion as well as
in the loop expansion up to one loop. After recognizing the ``dissipative
terms'' in these approximations,  we find that these cannot be simply replaced
by a term of the form $\Gamma \dot{\phi}(t)$.

In the case where
there is no symmetry breaking, we generate the effective Langevin
equation for the system, in the one-loop approximation, and we find a
multiplicative, {\em non}-Markovian kernel for the dissipational term as well
as a ``colored'' noise term. Thus this is very different than the simple
friction term described above.
In fact, it is easy to see that there is {\em no}
limit in which the dissipational dynamics we find can be described by a term
proportional to the time derivative of the field.

In the case in which a discrete symmetry is broken, so that there are no
Goldstone modes, the one-loop equation can be linearized in the amplitude of
oscillations about the non-zero expectation value of the field. Even in this
linearized approximation, while a term involving the first time derivative of
the field does appear, it is convolved with a non-trivial kernel, that again
has no limit (at zero temperature) in which it becomes a local term.

 We show that the loop expansion or the
amplitude expansion for the dynamics of the field expectation value {\em must}
break down at
long times. In the loop expansion, the order $\hbar$ term has an amplitude that
grows in time due to resonance phenomena induced by the existence of a two
particle threshold. Thus, it will eventually dominate over the tree amplitude,
and perturbation theory will then break down. Dissipation can only be
understood beyond perturbation theory. We then use the Hartree approximation
in the single scalar field case and the large $N$ approximation in an $O(N)$
symmetric theory to try to understand the effects of quantum fluctuations on
the oscillations of the field expectation value as well as particle production
due to these
same oscillations. The Hartree approximation reveals dissipative effects due to
particle production and open channels. However, we find that the damping is not
exponential, and asymptotically, the field expectation value undergoes undamped
oscillations with an amplitude that depends on the coupling and the initial
conditions. The same features are found in the large N limit in the O(N)
model in the case
of unbroken symmetry. Here we find that the asymptotic oscillatory behavior
can be  expressed in terms of elliptic functions.

The dissipation in the system is caused by a Landau damping\cite{landau} type
of process.
This is a collisionless process  which in
our case is associated with the particle production arising from
 oscillations of the field expectation value via parametric amplification.
These quantum fluctuations
 then react
back on the field expectation value, but {\em not} in phase with it, thus
damping the
oscillations. This behavior is similar to that found in the damping of strong
electric fields in a collisionless regime\cite{klug}.

An unexpected and quite remarkable result comes from the analysis of the $O(N)$
model in large $N$
when the symmetry is spontaneously broken. We find that the field expectation
value
oscillations are damped very quickly in this case, much more so than in the
unbroken situation. This by itself is perhaps not so unusual, since the
existence of Goldstone modes allows the field to lose energy by radiating a
large number of soft Goldstone particles. What is unusual, however, is that
there are some initial conditions, corresponding to ``slow-roll''
behavior\cite{kolbturner},
for which the field expectation value relaxes (via dissipation) to a value that
is very
close to the origin. This final value is a minimum of the effective action
for which the expectation
value of the scalar field has moved very slightly from the initial value.
Most of the potential energy is converted into long-wavelength Goldstone
bosons.
 Thus the symmetry breaking phase transition is
actually {\em dramatically modified} by the dissipational dynamics. These
initial
conditions are generically ``slow-roll'' conditions in which the scalar field
begins very close to the top of the potential hill and the couplings are
extremely weak (in our case about $10^{-7}$ or smaller).  This is one of the
most
striking results of this article.

We provide extensive numerical evidence for all our assertions; we can
follow the evolution of the field expectation value, the quantum fluctuations
and we also
compute the number of particles created via the method of Bogoliubov
coefficients.

In the next section we provide an outline of the formalism we use to generate
the equations for the expectation value of the scalar field that include the
effects of the fluctuations
of the quantum fluctuations. We then investigate the meaning of these equations
within the one-loop and the amplitude expansion. In section III and IV
we utilize
non-perturbative approximations such as the Hartree approximation and the large
$N$ approximation in the case of the $O(N)$ model to perform our numerical
calculations.
After this analysis, we reconcile the dissipative dynamics that we find
 with the concept of time reversal invariance.

Section V contains our conclusions, in which we
compare our results to other results on dissipation and reheating
obtained in the literature. Two appendices are
included which contain some of the technical details of our calculations.

\newpage

\section{Non-equilibrium field theory and equations of motion}

The generalization of statistical mechanics techniques to the description of
non-equilibrium processes in quantum field theory has been available for a long
time\cite{schwinger,mahan,keldysh,mills,zhou} but somehow has not yet been
accepted as an integral part of the available tools to study field theory in
extreme environments. We thus begin by presenting  a somewhat pedagogical
introduction to the subject for the non-practitioner.

The non-equilibrium description of a system is determined by the time evolution
of the density matrix that describes it.  This time evolution (in the
{Schr\"{o}edinger} picture) is determined by the quantum Liouville equation:

\begin{equation}
i\hbar \frac{\partial \rho(t)}{\partial t} = \left[H(t),\rho(t)\right]
\label{liouville1}
\end{equation}

A non-equilibrium situation arises whenever the Hamiltonian
does not commute with the density matrix. Here we allow for
an {\em explicitly} time dependent Hamiltonian, which might be the case
if the system is in an external time dependent background, for example.

The formal solution of the Liouville equation is:
\begin{equation}
\rho(t) = U(t,t_0)\rho(t_0)U^{-1}(t,t_0) \label{rhooft}
\end{equation}
with $\rho(t_0)$ the density matrix at some initial time $t_0$
that determines the initial condition for the  evolution.
Ensemble averages of operators and correlation functions
 are obtained as
\begin{eqnarray}
\langle {\cal{O}}(t) \rangle  & = &
Tr\left[{\cal{O}}U(t,t_0)\rho(t_0)U^{-1}(t,t_0)\right]/Tr\rho(t_0) \\
\langle {\cal{O}}(t_1){\cal{O}}(t_2) \rangle
                              & = &
Tr\left[{\cal{O}}U(t_1,t_2){\cal{O}}U(t_2,t_0)
\rho(t_0)U^{-1}(t_1,t_0)\right]/Tr\rho(t_0) \label{trace}
\end{eqnarray}

The expressions above have a very intuitive meaning. To compute the
ensemble average, take the
initial state (or density matrix), evolve it forward in time from
$t_0$ to $t$, insert the operator in question and evolve the state backwards
in time back to $t_0$. For the correlation function ($t_1>t_2$),
evolve the initial state to $t_2$, insert the operator, evolve it
further to $t_1$, insert the second operator, and finally evolve the state
backwards to $t_0$. In most cases of interest the initial density
matrix is either thermal or a pure state corresponding to the
ground state of some initial Hamiltonian. In both cases the
initial density matrix is of the form:
\begin{equation}
\rho(t_0) = e^{-\beta H_i};
\end{equation}
the ground state of $H_i$ can be projected out by taking the
limit $\beta \rightarrow \infty$. The system will undergo
non-equilibrium evolution
whenever the time evolution operator does not commute with
the initial Hamiltonian $H_i$. It is convenient to introduce
a time dependent Hamiltonian $H(t)$ such that $H(t)=H_i$ for
$-\infty \leq t \leq t_0$ and $H(t)=H_{evol}(t)$ for $t > t_0$, where
$H_{evol}(t)$ is the evolution Hamiltonian that determines the
dynamics. This corresponds to the assumption that the system
has been in equilibrium up to $t_0$ and will evolve out of
equilibrium thereafter. Real time {\it equilibrium} correlation
functions can be obtained if $H$ is constant in time for all
times.

Correlation functions and ensemble averages can be obtained by considering
evolution forward in time (from $t_0 \rightarrow -\infty$) insertion of
operators and backwards to the original time, and finally down the imaginary
time axis to $t_0- i\beta$ to account for the initial thermal condition. Taking
the trace  amounts to identifying the initial and final field configurations
and performing a functional integral over this configuration. Thus one is led
to consider path integrals in the complex time plane.

Correlation functions or operator averages may be obtained as
usual by coupling sources and taking functional derivatives with
respect to them.

 Thus we are led to
consider the following generating functional, in terms of time evolution
operators in presence of sources
\begin{equation}
Z[J^+, J^-, J^{\beta}]= Tr\left[ U(T-i\beta,T;J^{\beta})
U(T,T';J^-)U(T',T;J^+)\right]\label{generatingfunctional}
\end{equation}
with $T\rightarrow -\infty ; T'\rightarrow \infty$.
The denominator in (\ref{trace}) is simply $Z[0,0,0]$ and
may be obtained in a series expansion in the interaction by
considering $Z[0,0,J^{\beta}]$.
By inserting a complete set of field eigenstates between the
time evolution operators, finally, the generating functional
$Z[J^+,J^-,J^{\beta}]$ may be written as
\begin{eqnarray}
Z[J^+,J^-,J^{\beta}] & = & \int D \Phi D \Phi_1 D \Phi_2
\int {\cal{D}}\Phi^+ {\cal{D}}\Phi^-
{\cal{D}}\Phi^{\beta}e^{i\int_T^{T'}\left\{{\cal{L}}[\Phi^+,
J^+]-
{\cal{L}}[\Phi^-,J^-]\right\}}\times   \nonumber\\
                     &   &  e^{i\int_T^{T-
i\beta_i}{\cal{L}}[\Phi^{\beta}, J^{\beta}]}
\label{pathint}
\end{eqnarray}
with the boundary conditions $\Phi^+(T)=\Phi^{\beta}(T-
i\beta_i)=\Phi \;
; \; \Phi^+(T')=\Phi^-(T')=\Phi_2 \; ; \; \Phi^-
(T)=\Phi^{\beta}(T)=
\Phi_1$.
This may be recognized as a path integral along a contour
in complex time.
As usual the path integrals over the quadratic forms may be
evaluated and one obtains the final result for the partition
function
\begin{eqnarray}
& & Z[J^+,J^-, J^{\beta}] = \exp{\left\{i\int_{T}^{T'}dt
\left[{\cal{L}}_{int}(-i\delta/\delta J^+)-
{\cal{L}}_{int}(i\delta/\delta
J^-)\right] \right \}} \times \nonumber \\
& &
\exp{\left\{i\int_{T}^{T-
i\beta_i}dt{\cal{L}}_{int}(-i\delta/\delta J^{\beta})
\right\}} \exp{\left\{\frac{i}{2}\int_c
dt_1\int_c dt_2 J_c(t_1)J_c(t_2)G_c (t_1,t_2) \right\}},
 \label{partitionfunction}
\end{eqnarray}
where $J_c$ stands for the currents evaluated along the contour.
The $G_c$'s are the Green's functions on the
contour\cite{calzettahu,calzetta,niemi,jordan}(the spatial
arguments were suppressed).

In the limit $T \rightarrow -\infty$, the contributions from
the terms in which one of the currents is $J^+$ or $J^-$ and
the other is  $J^{\beta}$ vanish when computing correlation
functions in which the external legs are at finite {\it real
time}, as a consequence of the Riemann-Lebesgue lemma. For
this {\it real time} correlation functions, there is no
contribution from the $J^{\beta}$ terms that cancel between
numerator and denominator. Finite temperature enters through
the boundary conditions on the Green's functions. For the calculation of finite
{\it real time}
correlation functions, the generating functional simplifies
to\cite{calzetta,calzettahu}
\begin{eqnarray}
Z[J^+,J^-] & = & \exp{\left\{i\int_{T}^{T'}dt\left[
{\cal{L}}_{int}(-i\delta/\delta J^+)-
{\cal{L}}_{int}(i\delta/\delta
J^-)\right] \right \}} \times \nonumber \\
           &   & \exp{\left\{\frac{i}{2}\int_T^{T'}
dt_1\int_T^{T'} dt_2 J_a(t_1)J_b(t_2)G_{ab} (t_1,t_2)
\right\}}
 \label{generatingfunction}
\end{eqnarray}
with $a,b = +,-$.

  The Green's functions that enter in the integrals along
the  contours in equations (\ref{partitionfunction},
\ref{generatingfunction})
  are given by (see above references)
\begin{eqnarray}
\langle T \Phi(\vec{r}_1,t_1) \Phi(\vec{r}_2,t_2) \rangle & = & -i
G^{++}(\vec{r}_1,t_1;\vec{r}_2,t_2) \label{timeord} \\
\langle \tilde{T} \Phi(\vec{r}_1,t_1) \Phi(\vec{r}_2,t_2) \rangle
                                                          & = & -i
G^{--}(\vec{r}_1,t_1;\vec{r}_2,t_2) \label{antitimeord} \\
\langle  \Phi(\vec{r}_1,t_1) \Phi(\vec{r}_2,t_2) \rangle  & = &  i
G^{+-}(\vec{r}_1,t_1;\vec{r}_2,t_2) \label{plusmin}
\end{eqnarray}
with $T$ and $\tilde{T}$ the time ordering  and anti-time
ordering symbols.
The Green's functions above are written in terms of the homogeneous
solutions to the quadratic form of the Lagrangian (free fields) $G^>,\ G^<$
as
\begin{eqnarray}
G^{++}(\vec{r}_1,t_1;\vec{r}_2,t_2)  & = &
G^{>}(\vec{r}_1,t_1;\vec{r}_2,t_2)\Theta(t_1-t_2) +
G^{<}(\vec{r}_1,t_1;\vec{r}_2,t_2)\Theta(t_2-t_1)
\label{timeordered}\\
G^{--}(\vec{r}_1,t_1;\vec{r}_2,t_2)  & = &
G^{>}(\vec{r}_1,t_1;\vec{r}_2,t_2)\Theta(t_2-t_1) +
G^{<}(\vec{r}_1,t_1;\vec{r}_2,t_2)\Theta(t_1-t_2)
\label{antitimeordered} \\
G^{+-}(\vec{r}_1,t_1;\vec{r}_2,t_2)  & = & -
G^{<}(\vec{r}_1,t_1;\vec{r}_2,t_2) \label{plusminus}\\
G^{-+}(\vec{r}_1,t_1;\vec{r}_2,t_2)  & = & -
G^{>}(\vec{r}_1,t_1;\vec{r}_2,t_2) = -
G^{<}(\vec{r}_2,t_2;\vec{r}_1,t_1)
\label{minusplus}\\
G^{>}(\vec{r}_1,t_1;\vec{r}_2,t_2)   & = & i
\langle\Phi(\vec{r}_1,t_1)\Phi(\vec{r}_2,t_2)\rangle
\label{greater} \\
G^{<}(\vec{r}_1,T;\vec{r}_2,t_2)     & = &
G^{>}(\vec{r}_1,T-i\beta_i;\vec{r}_2,t_2)
\label{periodicity}
\end{eqnarray}
The condition (\ref{periodicity}) is recognized as the
periodicity condition in imaginary time and is a result of
considering an equilibrium situation for $t<t_0$.
The
functions $G^{>},\ G^{<}$, which are the homogeneous solutions of the
quadratic form, with appropriate boundary conditions,
will be constructed explicitly below.

The Feynman rules are the same as in regular perturbation
theory,  with the proviso that there are {\it two interaction vertices}
corresponding to the $\pm$ branches, with opposite signs (arising from the
complex conjugation of the unitary time evolution operator for evolution
backwards in time).

This formulation in terms of time evolution along a contour in complex
time has been used many times in non-equilibrium statistical mechanics.
 There are many clear articles in the literature
using this techniques to study real time correlation
functions\cite{calzettahu,calzetta,jordan,landsman,semenoffweiss,kobeskowalski}
 and effective actions out of
equilibrium\cite{paz,mazzi,marcelo,hubanff}.

This formulation has already been used by some of us previously to
study the dynamics of phase transitions\cite{boyveg}.

In our analysis we will take the $\beta \rightarrow \infty$
limit (zero temperature) following the argument provided in the
introduction. We will also limit ourselves to a Minkowski space-time rather
than  expanding FRW space-time by
considering phenomena whose time variation happens on scales much shorter than
the expansion time $H^{-1}$ of the universe, where $H$ is the Hubble parameter.

\subsection{Equations of motion:}

The method briefly discussed above allows the derivation of
the effective equations of motion for a coarse grained field.
To focus the discussion, consider the situation of a scalar
field theory with Lagrangian density
\begin{equation}
{\cal{L}} = \frac{1}{2}\partial_{\mu}\Phi\partial^{\mu}\Phi-
\frac{1}{2}m^2\Phi^2-\frac{\lambda}{4!}\Phi^4 \label{Lagrangian}
\end{equation}
The ``coarse grained'' field (order parameter) is defined
as
\begin{equation}
\phi(t) = \langle \Phi(x,t) \rangle=
\frac{Tr\Phi\rho(t)}{Tr\rho(t)} \label{ordpar}
\end{equation}
where we have used translational invariance. We write the
field as $\Phi(x,t) = \phi(t)+\psi(x,t)$ with $\psi(x,t)$
the fluctuations, obeying
 $\langle \psi(x,t) \rangle =0$
and consider $\phi(t)$ as a background field in the evolution
Hamiltonian. The non-equilibrium generating functional requires
\begin{equation}
{\cal{L}}[\phi+\psi^+]-{\cal{L}}[\phi+\psi^-]= \left\{
\frac{\delta {\cal{L}}}{\delta \phi}\psi^+ +
{\cal{L}}_0[\psi^+]-\lambda\left(\frac{\phi^2(\psi^+)^2}{4}+
\frac{\phi(\psi^+)^3}{6}+\frac{(\psi^+)^4}{4!}\right)\right\}-
\left\{\psi^+ \rightarrow \psi^-\right\} \label{plusminuslag}
\end{equation}
with ${\cal{L}}_0[\psi^{\pm}]$ the free field Lagrangian
density of a field of mass $m$. We can now ``integrate out'' the
fluctuations, thus obtaining the  non-equilibrium effective action for the
``coarse grained'' background field.

 The {\it linear}, cubic and quartic
terms in $\psi^{\pm}$ are treated as perturbations, while the terms
$\phi^2(t) (\psi^{\pm})^2$ may either be treated as perturbations,
if one wants to generate a perturbative expansion in terms of
the {\it amplitude} of the coarse grained field, or alternatively,
they may be absorbed into a time dependent mass term for the
fluctuations, if one wants to generate a {\it loop} expansion.
We will now study both cases in detail.

\subsection{Amplitude expansion}
\subsubsection{\bf Discrete Symmetry}

We treat the term $\phi^2(t) (\psi^{\pm})^2$ as perturbation,
along with the linear, cubic and quartic terms. The conditions
\begin{equation}
\langle \psi^{\pm}(x,t) \rangle = 0 \label{tadpole}
\end{equation}
will give rise to the effective non-equilibrium equations
of motion for the background field. This is the generalization
of the ``tadpole'' method\cite{weinberg} to non-equilibrium
field theory. Since the mass of the fluctuations is the bare
mass, the essential ingredient for the non-equilibrium propagators
introduced above are the spatial Fourier transforms of the homogeneous
solutions
to the free field quadratic forms:
\begin{eqnarray}
G^{>}_k(t_1,t_2) & = &  \frac{i}{2\omega_k}e^{-i \omega_k(t_1-t_2)}
\label{G>} \\
G^{<}_k(t_1,t_2) & = &  \frac{i}{2\omega_k}e^{i \omega_k(t_1-t_2)}
\label{G<} \\
\omega_k         & = & \sqrt{k^2+m^2}. \label{omegak}
\end{eqnarray}

To ${\cal{O}}(\phi^3)$ the diagrams are shown in figures (1.a-1.h).
Since the $(++)$ and the $(--)$ propagators are independent,
 the
diagrams (1.a-1.h) finally lead to the equation of motion

\begin{eqnarray}
\ddot{\phi}(t)+m^2 {\phi}(t)+\frac{\lambda}{6}{\phi}^3(t)
& + & \frac{\lambda}{2}{\phi}(t)\int \frac{d^3k}{(2\pi)^3}
\frac{1}{2\omega_k} \nonumber \\
& - & \frac{\lambda^2}{4}{\phi}(t)\int^t_{-\infty}dt'\phi^2(t')
\int \frac{d^3k}{(2\pi)^3}\frac{\sin[2\omega_k(t-t')]}{2\omega^2_k} = 0
\label{eqnofmotI}
\end{eqnarray}
Notice that the diagrams of figures (1.e-1.f) contain the
one-loop two-particle threshold typical of the  two particle to two
particle scattering amplitude. The last term in (\ref{eqnofmotI})
is the {\it real-time} expression for this one-loop contribution.
 We can perform an integration by parts in the time integral,
 discarding the contribution
at $t =-\infty$ by invoking an adiabatic switching-on
convergence factor to obtain
\begin{eqnarray}
&  & \ddot{\phi}(t)+m^2 {\phi}(t)+\frac{\lambda}{6}
{\phi}^3(t)+
 \frac{\lambda}{2}{\phi}(t)\int \frac{d^3k}{(2\pi)^3}
\frac{1}{2\omega_k} \nonumber \\
&- &
\frac{\lambda^2}{4}{\phi}^3(t)\int
\frac{d^3k}{(2\pi)^3} \frac{1}{4\omega^3_k}
 +  \frac{\lambda^2}{4}{\phi}(t)\int^t_{-\infty}dt'\phi(t')
\dot{\phi}(t')
\int \frac{d^3k}{(2\pi)^3}
\frac{\cos[2\omega_k(t-t')]}{2\omega^3_k} = 0.
\label{eqnofmotII}
\end{eqnarray}
The fourth term in the above equation is recognized as a
mass renormalization and the fifth term as the coupling
constant renormalization. Using a Fourier expansion for
$\phi(t)$ and its derivative, it is straightforward to see
that after integration in the time variable, the remaining
integration in $k$ is {\it ultraviolet finite}.
There are several noteworthy features of this expression that
point to a more complicated description of dissipative
processes in field theory. The first such feature is an expected one.
The ``dissipative'' contribution, that is, the last
term in the above equation, has a non-Markovian (i.e. memory-retaining)
kernel. Secondly, the equation is {\it non-linear} in the amplitude of
the coarse grained variable. These features are in striking
contradiction with a simple $\Gamma \dot{\phi}$ term in the
equation of motion. Originally\cite{abbottwise} such a term
was argued to arise if the scalar field in question is coupled
(linearly) to other fields in the theory and truly speaking
such a situation does not arise within the present context.
However the issue of a memory-kernel (non-Markovian) is quite
general\cite{morikawa}, and a legitimate question to ask is:
is there a Markovian (local) limit of this kernel? Such a limit
would imply that
\begin{equation}
{\cal{K}}(t-t') \approx D \delta(t-t') \label{markovian}
\end{equation}
with ${\cal{K}}(t-t')$ being the non-local kernel present in the last term of
(\ref{eqnofmotII}).
However, we find that
at small $(t-t')$ the kernel has typical logarithmic
divergences.

If a local approximation were valid,
the ``dissipative constant'' D may be
found by integrating the kernel in time. A straightforward
calculation shows that infrared divergences give a divergent
answer for such constant, clearly indicating that there is
no Markovian limit for this kernel.
In reference\cite{marcelo} a Markovian limit was argued to be
available in the high temperature limit; since we are working
at zero temperature the approximations invoked there do not
apply.
Thus, this lowest order calculation reveals two conclusive
features that will persist in higher orders and even in
non-perturbative calculations (see below): the ``dissipative
contribution '' to the equations of motion obtained by
integrating out the fluctuations are typically non-linear and, furthermore,
they
do not allow for a Markovian (local) description. We will
postpone a numerical analysis of these equations until we study
the full one-loop equations of motion below.

\subsubsection{Broken symmetry}

In the case of unbroken symmetry, the above equation of motion
does not admit a linearization of the dissipative contribution.
However, in the case when the symmetry is spontaneously broken such
a linearization is possible\cite{paz}.
In this case let $m^2=-\mu^2$ and
let us look for small oscillations around one of the minima.
To achieve this,
write $\phi(t)= \sqrt{6\mu^2/ \lambda} +\delta(t)$.
 The mass term for the fluctuations
now becomes $2\mu^2$, and we will only keep the linear terms in
$\delta$ in (\ref{plusminuslag}). We now follow the steps
leading to the equation of motion obtained before. That is,
$\langle \Psi^{\pm} \rangle = 0$ is imposed to one-loop order. After
integration
by parts in the time integral (again with an adiabatic switching-on
convergence factor) there appear
several tadpole contributions. Those that are independent of $\delta$
renormalize  $\phi_0$ corresponding to a shift of the position of the vacuum
expectation value, while those that are linear in $\delta$ renormalize
the mass. We thus obtain in the linearized approximation:
\begin{equation}
\ddot{\delta}+2\mu_R^2 \delta+\frac{3\mu_R^2}{2}\lambda_R
\int_{-\infty}^t\dot{\delta}(t')\int \frac{d^3k}{(2\pi)^3}
\frac{\cos[2\tilde{\omega}_k(t-t')}{2\tilde{\omega}^3_k}=0
\label{lineareqofmotion}
\end{equation}
with $\tilde{\omega}_k=\sqrt{k^2+2\mu^2}$. This expression is
similar to that found by Paz\cite{paz}. In that reference the
 {\it short time}
behavior was analyzed.  In order to solve this
equation for all times, we must specify an initial condition.
We will {\it assume} that
for $t<0$ $\delta(t<0)=\delta_i$ and $\dot{\delta}(t<0)=0$ and that equation
(\ref{lineareqofmotion}) holds for $t>0$\cite{note}.

Under this assumption, the linearized equation can be solved via the Laplace
transform. This solution corresponds to summing Dyson's series for the
propagator with the one-loop contributions depicted in figure (1) (for the
linearized case). The Laplace transform exhibits the
 two-particle cut\cite{calzettahu, paz} in the
analytic continuation of the transform variable $s$.
However the inverse Laplace transform proves to be
extremely difficult to carry out. We will content ourselves
at this stage with
a perturbative solution to ${\cal{O}}(\lambda)$. Thus we write
\begin{equation}
\delta(t)=\delta_i \cos[\sqrt{2}\mu_R t]+\lambda
 \delta_1(t) \label{pert}
\end{equation}
The equation is then solved order by order in $\lambda$. The corresponding
equation for $\delta_1(t)$ is now easily solved via the Laplace transform. The
transform exhibits resonances at $\pm \sqrt{2}\mu_R$ and the large time
behavior is dominated by the secular term
$\delta_1(t) \approx t\sin[\sqrt{2}\mu_R t]$.
Thus the conclusion from the
linearized approximation is that it is valid only for {\it short times}; the
long time behavior will not be captured by this perturbative linearized
approximation. This conclusion will be confirmed with a numerical analysis of
the full
one-loop case below.
 The analysis provided in
references\cite{calzettahu, paz, mazzi} in which a dissipative
behavior was observed is not quite consistent, because their solution contains
higher order terms in the coupling that are not warranted in the approximation
considered. Furthermore, even if the implied resummation is accepted the
solution found by these authors
is only valid at very short times, corresponding to the region in the s-plane
(Laplace transform variable) far away from the resonances.
The physics of this effect is clear if analyzed in terms of
the Fourier transform. At short times the contribution
of large frequencies is substantial and the two-particle threshold is
available for dissipation. However at long times only low
frequencies have an important contribution, the two-particle
threshold is not available and the resonances below threshold
dominate giving a growing amplitude.

Furthermore, even
in this case in which there is an explicit linear velocity dependence in the
``dissipative kernel'', the same argument presented before applies.
There is no Markovian (local) limit in which this term
becomes simply $\Gamma \dot{\delta}$,
despite the fact that $\delta$ has a {\it linear coupling}
to the fluctuations.

\subsubsection{Continuum symmetries and Goldstone bosons}
An important issue that we want to study in detail in this
article is the process of dissipation by Goldstone bosons.
As discussed in the introduction, dissipation via particle
production becomes effective whenever the transferred energies
become larger than multiparticle thresholds. To lowest order,
as shown in the perturbative calculation of the previous
section, the lowest threshold  is the two-particle one, at an energy
twice the mass of the particle. In the
case of a spontaneously broken, continuous symmetry, there will be Goldstone
bosons and the thresholds are at zero energy. In this case any
amount of energy transfer may be dissipated by the Goldstone
modes.

This effect  may be studied in the
$O(2)$ linear sigma model with Lagrangian density
\begin{equation}
{\cal{L}} = \frac{1}{2}\partial_{\mu}\sigma\partial^{\mu}\sigma
+\frac{1}{2}\partial_{\mu}\pi\partial^{\mu}\pi
+
\frac{1}{2}\mu^2(\sigma^2+\pi^2)-\frac{\lambda}{4!}
(\sigma^2+\pi^2)^2 \label{Lagrangiansigma}
\end{equation}

We now write
\begin{equation}
\sigma(x,t)= \sqrt{\frac{6\mu^2}{\lambda}}+\delta(t)+\chi(x,t)
\label{fieldsplit}
\end{equation}
and use the tadpole method to impose:
\begin{equation}
\langle \pi(x,t) \rangle   =  0  \; \; ; \; \;
\langle \chi(x,t) \rangle  =  0 \label{chizero}
\end{equation}
Carrying out the same analysis in terms of Feynman diagrams to
linear order in $\delta(t)$ we find the following equation
of motion (after integrating by parts invoking an adiabatic
switching on convergence factor and absorbing the local terms
in proper renormalizations)
\begin{equation}
\ddot{\delta}+2\mu_R^2+6\mu_R^2 \lambda_R \int_{-\infty}^t dt'
\dot{\delta}(t')\left\{\frac{1}{4}{\cal{K}}_{\chi}(t-t')
+\frac{1}{36}{\cal{K}}_{\pi}(t-t')\right\}=0
\end{equation}
with ${\cal{K}}_{\chi}(t-t')$ the same kernel as in equation
(\ref{lineareqofmotion}) and
\begin{equation}
{\cal{K}}_{\pi}(t-t') = \int \frac{d^3k}{(2\pi)^3}
\frac{\cos[2|k|(t-t')]}{2|k|^3} \approx \ln[M(t-t')]
\end{equation}
where $M$ is an infrared cutoff introduced to define the
integral. This infrared divergence is the result of the fact
that the
threshold for Goldstone bosons is at zero energy-momentum.
This result clearly reflects many important features of
``dissipation'' via Goldstone bosons. First, as before, the
dissipative kernel cannot be described in a local (Markovian)
approximation even in this linearized theory in which the
arguments leading to a $\Gamma \dot{\delta}$ term would be valid.
Second and perhaps more important, the long time behavior
is clearly {\it beyond perturbation theory} as the contribution
from the non-local kernels is unbounded as a result of
 infrared divergences associated with Goldstone bosons.

\subsection{One-loop equations}

The full one-loop equations of motion are obtained by
absorbing the terms $\phi^2(t)(\psi^{\pm})^2$
in eq.(\ref{plusminuslag}) as a time dependent mass term for the
fluctuating fields. In order to determine the dynamics for the
background field, we will assume that $\phi(t<0)=\phi_0$ and that
at $t=0$ the background field is ``released'' with zero velocity.
Now we must find the corresponding non-equilibrium
Green's functions. Consider the following homogeneous solutions
of the quadratic form for the fluctuations
\begin{eqnarray}
\left[\frac{d^2}{dt^2}+\vec{k}^2+m^2+\frac{\lambda}{2}\phi^2(t)
\right]U^{\pm}_k(t) & = & 0 \label{modes} \\
U^{\pm}_k(0)= 1 \; \; ; \; \;
\dot{U}^{\pm}_k(0)  & = & \mp i \omega^0_k
\label{bc}
\end{eqnarray}
with
\begin{equation}
 \omega^0_k          =  \left[\vec{k}^2+m^2+
\frac{\lambda}{2}\phi^2(0)\right]^{\frac{1}{2}}. \label{freqini}
\end{equation}
 The boundary conditions (\ref{bc}) correspond to positive
$U^+$ and negative $U^-$ frequency modes for $t<0$
(the Wronskian of
these solutions is $2i \omega^0_k$).
Notice that $U_k^-(t)=[U_k^+(t)]^*$.
 In terms of these mode functions we obtain
\begin{equation}
G^>(t_1,t_2) = \frac{i}{2\omega^0_k}U_k^+(t_1)U_k^-(t_2)
\end{equation}
Now the equation of motion to one-loop is obtained from the
diagrams in figures (1.a), (1.d) (but with the modified propagator including
the contribution of $\phi(t)$). Thus to one-loop we find
the following equations
\begin{eqnarray}
\ddot{\phi}(t)+m^2 {\phi}(t)+\frac{\lambda}{6}{\phi}^3(t)+
\frac{\lambda \hbar}{2}{\phi}(t)\int \frac{d^3k}{(2\pi)^3}
\frac{|U^+_k(t)|^2}{2\omega^0_k} & = & 0 \label{oneloopeqn} \\
 \left[\frac{d^2}{dt^2}+\vec{k}^2+m^2+\frac{\lambda}{2}
\phi^2(t)
\right]U^+_k(t)                  & = & 0 \label{modeeqn} \\
U^{+}_k(0)= 1 \; \; ; \; \;
\dot{U}^{+}_k(0)                 & = & -i \omega^0_k
\label{bceqn}
\end{eqnarray}
where we have restored the $\hbar$ to make the quantum
corrections explicit.
This set of equations  clearly shows how the expectation
value (coarse grained variable) ``transfers energy'' to the mode
 functions via a time dependent frequency,
 which then in turn modify the equations of motion
for the expectation value. The equation for the mode functions,
 (\ref{modeeqn}) may be solved in a perturbative expansion in terms
of $\lambda \phi^2(t)$ involving the {\em retarded}
 Green's function. To first order in that expansion one
recovers eq.(\ref{eqnofmotI}).

 Before attempting a numerical solution of the
above equations it is important to understand the renormalization
aspects. For this we need the large $k$ behaviour of the mode
functions which is obtained via a WKB expansion as in references
\cite{boyveg,boyholveg} and to which the reader is referred to
for details. We obtain
\begin{equation}
\int \frac{d^3k}{(2\pi)^3}
\frac{|U^+_k(t)|^2}{2\omega^0_k} = \frac{\Lambda^2}{8\pi}-
\frac{1}{8\pi}\left[m^2+\frac{\lambda}{2}{\phi}^2(t)\right]
\ln\left[\frac{\Lambda}{\kappa}\right]+ \mbox{ finite }
\label{divergences}
\end{equation}
where $\Lambda$ is an ultraviolet cutoff and $\kappa$ an
arbitrary renormalization scale. From the above expression
it is clear how the mass and coupling constant are renormalized.
It proves more convenient to subtract the one-loop contribution at
$t=0$ and absorb a finite renormalization in the mass, finally
 obtaining
the renormalized equation of motion
\begin{equation}
\ddot{\phi}(t)+m_R^2 {\phi}(t)+\frac{\lambda_R}{6}{\phi}^3(t)+
 \frac{\lambda_R \hbar}{8\pi^2}{\phi}(t)\int^{\Lambda}_0 k^2 dk
\frac{\left[|U^+_k(t)|^2-1\right]}{\omega^0_k} +
\frac{\lambda^2_R \hbar}{32\pi^2}
{\phi}(t)(\phi^2(t)-\phi^2(0))\ln\left[\Lambda / \kappa\right]
 =  0 \label{renormeqnofmot}
\end{equation}
In the equations for the mode functions the mass and coupling
may be replaced by the renormalized quantities to this order.
One would be tempted to pursue a numerical solution of these
coupled equations. However doing so would not be consistent,
since these equations were obtained only to order $\hbar$
and a naive numerical solution will produce higher powers of
$\hbar$ that are not justified.

 Within the spirit of the loop
expansion we must be consistent and only keep terms of order
 $\hbar$. First we introduce  dimensionless
variables
\begin{equation}
\eta(t)    =  \sqrt{\frac{\lambda_R}{6 m^2_R}} \phi(\tau)
\; \; ; \; \; \tau        =   m_R t \; \; ; \; \;
q        =  \frac{k}{m_R} \; \; ; \; \;
g = \frac{\lambda_R \hbar}{8\pi^2} \label{dimensionlessquantities}
\end{equation}
and expand the field in terms of
$g$ as
\begin{equation}
\eta(\tau)  =  \eta_{cl}(\tau)+g \eta_1(\tau)+ \cdots \label{fieldexpansion} \\
 \end{equation}
Now the equations of motion consistent up to ${\cal{O}}(\hbar)$
become
\begin{eqnarray}
&&\ddot \eta_{cl}(\tau)+ \eta_{cl}(\tau)+\eta^3_{cl}(\tau)    =  0
\label{classpart} \\
&&\ddot{\eta}_1(\tau)+\eta_1(\tau)+3\eta^2_{cl}(\tau)
                                      \eta_1(\tau) +
\eta_{cl}(\tau)
\int^{\Lambda/m_R}_0 q^2 dq
\frac{\left[|U^+_q(\tau)|^2-1\right]}
{\left[q^2+1+3\eta^2_{cl}(0)\right]^{1/2}}         + \nonumber \\
&&\frac{3}{2}\eta_{cl}(\tau)\left(\eta^2_{cl}(\tau)-\eta^2_{cl}(0)\right)
\ln[\Lambda/m_R]                                     =  0  \label{quancorr}
\end{eqnarray}

The solution to equation (\ref{classpart}) is an elliptic function.
The equations for the mode functions become
\begin{equation}
\left[\frac{d^2}{d\tau^2}+q^2+1+3\eta^2_{cl}(\tau)\right]U^+_q(\tau)
                                                   =  0 \label{dimenmods}
\end{equation}
with the boundary conditions as in eq.(\ref{bceqn}) in terms of the
dimensionless frequencies and
where for simplicity, we have chosen the renormalization scale $\kappa=m_R$.
The chosen
boundary conditions $\eta(0) = \eta_0 \; ; \; \dot{\eta}(0)=0$
can be implemented as
\begin{equation}
\eta_{cl}(0) = \eta_0 \; \; ; \; \; \dot{\eta}_{cl}(0)=0 \; \; ; \; \;
\eta_1(0) =0 \; \; ; \; \; \dot{\eta}_1(0) = 0 \label{eqnsplit}
\end{equation}
In fig.(2) we show $\eta_1(\tau)$ with the above boundary
conditions with $\eta_0=1$ and $g=0.1$. A cutoff $\Lambda /m_R =100$ was
chosen but no cut-off sensitivity was detected by varying the
cutoff by a factor 3. Notice that the amplitude grows as
a function of time. This phenomenon  can be understood as
follows.  A first hint was obtained in the case of the linearized
equations for a broken discrete symmetry.
There we learned
that because of resonances below the two-particle threshold, the
amplitude grows at long times. This is  the behavior shown in
fig.(2).  An alternative and perhaps more convincing argument
is the following. The mode functions $U^+_q(\tau)$ obey a
{Schr\"{o}edinger}-like
equation with a potential that is a {\em periodic} function
of time because the classical solution is periodic. Let us call the
period of the classical solution $T$. Floquet's
theorem\cite{matthews} guarantees the existence of  solutions that obey
\begin{equation}
U^+_q(\tau+T)=e^{\mu T}U^+_q(\tau) \label{floquet}
\end{equation}
The Floquet indices $\mu$ are functions of the parameters of the
potential. The classical solution is an elliptic function, and in this case
the {Schr\"{o}edinger} equation for the modes may be shown to be
a Lam\'{e} equation with $n=2$ (see ref.\cite{ince}) whose solutions are
Weierstrass functions. That is, a two-zone potential, with two forbidden
and two  allowed bands for $q>0$.
The Floquet indices $\mu$ are pure imaginary for large
$q$, but they are real and positive for $q$ near zero. That is, the long
wavelengths
belong to an unstable band.

A more intuitive understanding at a simpler level
ensues if we look at small oscillations near the origin for
the classical solution $\eta_{cl}(\tau) \approx \eta_{cl}(0)
\cos(\tau)$. Then the {Schr\"{o}edinger} equation for the modes becomes
a Mathieu equation\cite{matthews}, for which the dependence of
the Floquet indices on the parameters ($q\; ; \; \eta_{cl}(0)$) is
known\cite{matthews}. There are unstable bands in which the
Floquet index has positive real part for certain values of
these parameters.

Since the one-loop correction involves an
integral over  all wave-vectors, the values of $q$ in these
bands give growing contributions to the one-loop integral.

These instabilities of the one-loop
equations preclude a perturbative analysis of the process of
dissipation.

\subsubsection{Goldstone bosons at one-loop}

It proves illuminating to study the one-loop contribution to
the equations of motion for the scalar field expectation
value from Goldstone bosons. We proceed as in the previous
case but now with the Lagrangian density of the O(2) linear
sigma model (\ref{Lagrangiansigma}).
We now write
\begin{eqnarray}
\sigma(x,t) & = &  \sigma_0(t)+\chi(x,t)  \label{sigmasplit} \\
\langle \pi(x,t)\rangle
            & = & 0 \; \; ; \; \; \langle \chi(x,t)\rangle = 0 \label{zeroexp}
\end{eqnarray}
where  the terms $\chi^2 \sigma^2_0 \; ; \; \pi^2 \sigma^2_0$ are  now
included with the mass terms for the respective fields.
The procedure is exactly the same as in the previous case, but
we now use $\mu_R$ as the mass scale and introduce the dimensionless
variables of eq.(\ref{dimensionlessquantities}) in terms of this
scale. Performing the proper renormalizations and a subtraction
at $t=0$ we finally find the following equations for this
case
\begin{eqnarray}
\frac{d^2}{d\tau^2} \eta-\eta & + & \eta^3  +  g\eta\left\{
\int_0^{\Lambda/\mu_R}\frac{q^2}{W_{\chi}(q)}\left[|U^+_q(\tau)|^2-1\right]
+\frac{3}{2}\ln[\Lambda/\mu_R](\eta^2(\tau)-\eta^2(0))\right\} \nonumber \\
                              & + &
\frac{g}{3}\eta\left\{
\int_0^{\Lambda/\mu_R}\frac{q^2}{W_{\pi}(q)}\left[|V^+_q(\tau)|^2-1\right]
+ \frac{1}{2}\ln[\Lambda/\mu_R](\eta^2(\tau)-\eta^2(0))\right\}=0
\label{goldonelup}
\end{eqnarray}

The mode functions satisfy the equations
\begin{eqnarray}
\left[\frac{d^2}{d\tau^2}+q^2-1+3\eta^2(\tau)\right]U^+_q(\tau)
 & = & 0 \label{modechi} \\
\left[\frac{d^2}{d\tau^2}+q^2-1+\eta^2(\tau)\right]V^+_q(\tau)
 & = & 0 \label{modepi}
\end{eqnarray}
where the boundary conditions are the same as in (\ref{modeeqn})
but in terms of the initial frequencies $W_{\chi}(q)\; ; \;
W_{\pi}(q)$. As argued previously, for consistency we have to
expand the field as in eq.(\ref{fieldexpansion}) and keeping
only $\eta_{cl}$ in the equation for the modes. Notice that
whereas the minimum of the tree level potential corresponds to
$\eta^2_{cl}=1$ and the stable region for the $U$ modes is
$\eta^2_{cl} > 1/3$, the stable region for the $V$ modes (pion field)
is for $\eta^2_{cl} > 1$.
These instabilities require the choice of special initial conditions,
in particular for
 the initial frequencies. We choose
\begin{eqnarray}
W_{\chi}(q)= \sqrt{q^2+1+3\eta^2(0)} \label{freqchi} \\
W_{\pi}(q)= \sqrt{q^2+1+\eta^2(0)} \label{freqpi}
\end{eqnarray}
This choice corresponds to a gaussian initial state centered
at $\eta(0)$ but with {\em positive} frequencies at
 $\eta(0)=0$. This state evolves in a broken symmetry potential
from the position determined by $\eta(0)$. Alternatively, one
can think of this situation as changing the sign of the mass at
time $t=0$ from positive to negative.
For small oscillations of $\eta_{cl}$ around the
minimum at $\eta_{min}=1$, every time that the
classical field oscillates to the left of the minimum the
``pion'' field becomes unstable and grows, and thus its contribution to
the fluctuations becomes large. This behavior is depicted in
fig.(3) which shows $\eta_1(\tau)$ with the boundary conditions
as in (\ref{eqnsplit}) and the initial frequencies
 (\ref{freqchi},\ref{freqpi}) for the
mode functions where chosen with $\eta_{cl}(0)=0.6$
 and $\Lambda/ \mu_R =100$ but no cutoff
sensitivity was detected by increasing the cutoff by a factor of
2. This situation clearly exhibits Floquet indices with a
positive real part, and $\eta_{cl}(\tau)$ is periodic for small oscillations
around $\eta_{cl}=1$. The unstable wave vectors form a band
when $\eta_{cl}$ oscillates around this value. In each period
of $\eta_{cl}$ the amplitude grows because of these unstable
modes.

This figure clearly shows a dramatic growth in the amplitude
of the quantum correction at long times as a consequence of
the instabilities associated with the Goldstone mode.
As a consequence of this, perturbation theory {\em must} fail at long times.

\subsubsection{The Langevin equation}

A fundamental aspect of dissipation is that of decoherence which
plays an important role in studies of quantum
cosmology\cite{hubanff}. Furthermore
dissipation and fluctuations are related
by the fluctuation-dissipation theorem. This point has been
stressed by Hu and collaborators\cite{calhu}. The relation between
dissipative kernels and fluctuations and correlations of bath degrees
of freedom is best captured in a Langevin equation.

In this section we present a derivation of the corresponding
Langevin equation to one-loop order in an amplitude expansion.
The first step towards deriving the Langevin equation is to
determine the ``system'' and ``bath'' variables, once this
is done, one integrates out the ``bath'' variables obtaining
a (non-local) influence functional\cite{hubanff,feyn,legg} for the
system variables.

Starting with the generating functional of  eq.(\ref{pathint})
(for the case of zero temperature $\beta \rightarrow \infty$)
 we
separate the zero mode ($\phi^{\pm}$) from the field as
\begin{equation}
\Phi^{\pm}  =   \phi^{\pm}+ \psi^{\pm} \; \; ; \; \;
\int d^3x \psi^{\pm}(x,t)
            =  0 \label{knonzero}
\end{equation}
and consider $\phi$ as the ``system'' and $\psi$ as the ``bath''
to be integrated out. Because of the condition in eq.(\ref{knonzero}),
there are no terms linear in $\psi^{\pm}$ in the expansion of
the action in terms of these fields. Thus
\begin{eqnarray}
S[\phi^+,\psi^+]-S[\phi^-,\psi^-] & = & \Omega \left({\cal{S}}[\phi^+]-
{\cal{S}}[\phi^-]\right)+S_0[\psi^+]-S_0[\psi^-] \nonumber \\
                                  & - & \frac{\lambda}{4}
\left[(\phi^+(t))^2 \int d^3x (\psi^+(x,t))^2-
(\phi^-(t))^2\int d^3x (\psi^-(x,t))^2\right] \nonumber \\
                                  & + &
{\cal{O}}((\psi^{\pm})^3,(\psi^{\pm})^4)
\label{actionplumi}
\end{eqnarray}
where $\Omega$ is the spatial volume, ${\cal{S}}$ the action
per unit spatial volume, $S_0$ the free field
action   and the terms
 ${\cal{O}}((\psi^{\pm})^3,(\psi^{\pm})^4)$ will contribute at the
two-loop level and beyond. To this order, the coupling between
``bath'' and ``system'' is similar to the bi-quadratic coupling
considered in reference\cite{huzanpaz} (see also\cite{hubanff}).
Integrating out the $\psi^{\pm}$
fields in a consistent loop expansion gives rise to the
influence functional\cite{hubanff,feyn,legg} for the zero modes.
The  one loop diagrams
contributing to this functional up to ${\cal{O}}((\phi^{\pm})^4)$ are shown in
fig.(4).
In order to obtain the Langevin equation it is convenient
to introduce the center of mass ($\phi(t)$) and
relative ($R(t)$) coordinates
(these are the coordinates used in the Wigner transform of the
coordinate density matrix) as\cite{calhu,legg,schmid}
\begin{equation}
\phi^{\pm}(t) = \phi(t)\pm \frac{R(t)}{2} \label{CMcoord}
\end{equation}
 Using the Green's functions in eqs.(\ref{timeord}-\ref{greater})
with (\ref{G>},\ref{G<}) and being patient with the algebra
we find the effective action per unit spatial volume

\newpage

\begin{eqnarray}
{\cal{S}}_{eff}[\phi,R] & = & \int_{-\infty}^{\infty}dt\left\{
{\cal{L}}[\phi+R/2]-{\cal{L}}[\phi-R/2]-\frac{\lambda}{2}R(t)\phi(t)
\int\frac{d^3k}{(2\pi)^3}\frac{1}{2\omega_k} \right. \nonumber \\
                        & + & \left. \frac{\lambda^2}{4}
R(t)\phi(t) \int_{-\infty}^t dt' \phi^2(t')
\int\frac{d^3k}{(2\pi)^3}\frac{\sin[2\omega_k(t-t')]}{2\omega^2_k} \right \}
\nonumber \\
                        & + & i \frac{\lambda^2}{8}
 \int^{\infty}_{-\infty} dt \int^{\infty}_{-\infty}dt' R(t)R(t')\phi(t)\phi(t')
\int\frac{d^3k}{(2\pi)^3}\frac{\cos[2\omega_k(t-t')]}{2\omega^2_k} \nonumber \\
                        & + &
{\cal{O}}(R^3;R^4..). \label{effact}
\end{eqnarray}
The higher order terms ${\cal{O}}(R^3;...)$ receive contributions from two
and higher loops and give higher order corrections to the lowest order
(one-loop)
Langevin equation.
The imaginary part of the effective action above (last non-local term)
gives a contribution to the path integral that may be written in terms of
a stochastic field as
\begin{eqnarray}
&&\exp\left[-\frac{1}{2}\int^{\infty}_{-\infty} dt \int^{\infty}_{-\infty}dt'
R(t) {\cal{K}}(t,t')R(t')\right] \propto \int {\cal{D}}\xi
{\cal{P}}[\xi] \exp\left[i \int^{\infty}_{-\infty} dt \xi(t)R(t)\right]
 \label{noisefunctional} \\
&& {\cal{P}}[\xi] = \exp\left[-\frac{1}{2}\int^{\infty}_{-\infty} dt
\int^{\infty}_{-\infty}dt'
\xi(t) {\cal{K}}^{-1}(t,t')\xi(t')\right] \label{prob}
\end{eqnarray}
with
\begin{equation}
{\cal{K}}(t,t') = \frac{\lambda^2}{4} \phi(t)\phi(t')
\int\frac{d^3k}{(2\pi)^3}\frac{\cos[2\omega_k(t-t')]}{2\omega^2_k}\label{kernel}
\end{equation}
The non-equilibrium path integral now becomes (keeping track of volume factors)
\begin{equation}
Z \propto \int{\cal{D}}\phi {\cal{D}}R {\cal{D}}\xi {\cal{P}}[\xi]
\exp\left\{i \Omega  \left[{\cal{S}}_{reff}[\phi,R]+\int dt \xi(t)
R(t)\right]\right\}
\label{noisy}
\end{equation}
with ${\cal{S}}_{reff}[\phi,R]$ the real part of the effective action in
(\ref{effact}) and
with ${\cal{P}}[\xi]$ the gaussian probability distribution for the stochastic
noise variable given in (\ref{noisefunctional}).

The Langevin equation is obtained via the saddle point
condition\cite{hubanff,calhu,legg,schmid}
\begin{equation}
\frac{\delta {\cal{S}}_{reff}}{\delta R(t)}|_{R=0} = \xi(t)  \label{lange}
\end{equation}
leading to
\begin{eqnarray}
\ddot{\phi}(t)+m^2 {\phi}(t)+\frac{\lambda}{6}{\phi}^3(t)
& + & \frac{\lambda}{2}{\phi}(t)\int \frac{d^3k}{(2\pi)^3}
\frac{1}{2\omega_k} \nonumber \\
& - & \frac{\lambda^2}{4}{\phi}(t)\int^t_{-\infty}dt'\phi^2(t')
\int \frac{d^3k}{(2\pi)^3}\frac{\sin[2\omega_k(t-t')]}{2\omega^2_k} = \xi(t)
\label{langefin}
\end{eqnarray}
where the stochastic noise variable $\xi(t)$ has gaussian correlations
\begin{equation}
<<\xi(t)>>=0 \; ; \; <<\xi(t)\xi(t')>>=
{\cal{K}}(t,t') = \frac{\lambda^2}{4} \phi(t)\phi(t')
\int\frac{d^3k}{(2\pi)^3}\frac{\cos[2\omega_k(t-t')]}{2\omega^2_k}
\label{noisecorrel}
\end{equation}
Here, the double brackets stand for averages with respect to the gaussian
probability distribution ${\cal{P}}[\xi]$. We can see that the noise is
colored (not delta function correlated) and {\em multiplicative}. By
integrating by parts the ``dissipative kernel'' in eq.(\ref{langefin}) (the
last
non-local term) in the same way as done in eq.(\ref{eqnofmotII}), we can
clearly see that the resulting ``dissipative kernel'' and the noise correlation
function obey a generalized fluctuation-dissipation
theorem\cite{hubanff,calhu}. In the broken symmetry case, in the {\em
linearized approximation} (around the tree level minimum) and {\em if} the
k-integral could be replaced by a delta function, we would obtain the usual
fluctuation-dissipation relation. By taking the average over the stochastic
noise of (\ref{langefin}) with the gaussian probability distribution
(\ref{prob}) one obtains the equation of motion (\ref{eqnofmotII}) by replacing
the average of product of fields by the product of the averages (thus
considering the field as a classical background).

The higher order terms
in the effective action (influence functional) give rise to modifications to
the noise correlations, making them non-gaussian and involving more powers of
$\phi$ in the kernels, in principle these corrections may be computed
systematically
in a loop expansion.

Although this Langevin equation clearly exhibits the generalized
fluctuation-dissipation theorem connecting the ``dissipative'' kernel to the
correlations of the stochastic noise, it is a hopeless tool for any
evaluation of the dynamics. The long range kernels and the multiplicative
nature
of the noise prevent this Langevin equation from becoming a useful
tool. Its importance resides at the fundamental level in that it
 provides a direct link between fluctuation and dissipation including all
the memory effects and multiplicative aspects of the noise correlation
functions. This last correlation function is related to the decoherence
functional\cite{hubanff}.

\subsubsection{Failure of perturbation theory to describe dissipation}
This section has been devoted to a perturbative analysis of the
``dissipative aspects'' of the equation of motion for the scalar field.
Perturbation theory has been carried out as an amplitude expansion
and also up to  ${\cal{O}}(\hbar)$, both in the broken and
the unbroken symmetry case. In both cases
 we found both
analytically and numerically that the amplitude of the quantum
corrections {\em grow as a function of time}
and that the long time behavior cannot be captured in perturbation
theory.
This failure of perturbation theory to describe dissipation is
clearly understood from a very elementary but yet illuminating example:
the damped harmonic oscillator. Consider a damped harmonic oscillator
\begin{equation}
\ddot{q}+\Gamma \dot{q}+q=0 \label{dho}
\end{equation}
with $\Gamma \sim {\cal{O}}(\lambda)$ where $\lambda$ is a
small perturbative coupling. One can attempt to solve eq.(\ref{dho})
in a perturbative expansion in $\Gamma$. That is, set
$q(t)=q_0(t)+\Gamma q_1(t)+\cdots$.
The solution for $q_1(t)$ may be found by the Laplace transform:
\begin{equation}
q_1(t) \propto \Gamma t \cos(t) \label{q1}
\end{equation}
clearly exhibiting resonant behavior. This is recognized as a secular term.
If eq.(\ref{dho})
had been obtained as an effective equation of motion in a
{\em perturbative expansion} in $\lambda$, this would be the
consistent manner to solve this equation. However, we would be led
to conclude that perturbation theory breaks down at long times.
The correct solution is
\begin{equation}
q(t)= e^{-\frac{\Gamma}{2}t}\cos[\omega(\Gamma^2)t] \approx
\cos(t)-\frac{\Gamma}{2}t \cos(t) + {\cal{O}}(\Gamma^2).
\end{equation}
We see that the first order correction in $\Gamma$ is correctly
reproduced by perturbation theory, but in order to find
appreciable damping, we must wait a time $\sim {\cal{O}}(1/\Gamma)$
at which perturbation theory becomes unreliable.

In order to properly describe dissipation and damping one must
{\em resum} the perturbative expansion. One could in principle
keep the first order correction and exponentiate it in an
{\em ad-hoc} manner with the hope that this would be the correct
resummation. Although ultimately this {\em may be} the correct
procedure, it is by no means warranted in a field-theoretic
perturbative expansion, since in field theory, dissipation is
related to particle production and open multiparticle channels,
both very subtle and non-linear mechanisms.
Another hint that points to a resummation of the perturbative
series is provided by the set of equations
eqs.(\ref{classpart}-\ref{dimenmods}).
In eq.(\ref{classpart}), the classical solution is a periodic
function of time of constant amplitude, since the classical equation has
a conserved energy. As a consequence, the ``potential'' in the
equation for the modes (eq.(\ref{dimenmods})) is a periodic function of
time with constant amplitude. Thus although the fluctuations
react back on the coarse grained field, only the classically
conserved part of the motion of the coarse grained field enters
in the evolution equations of the mode functions. This is a
result of being consistent with the loop expansion, but clearly
this approximation is {\em not} energy conserving.

As we will point out in the next section, in an energy conserving scheme
the fluctuations and amplitudes will grow up to a maximum value and then
will always remain bounded at all times.

Thus in summary for this section, we draw the conclusion that
perturbation theory is not sufficient (without major ad-hoc assumptions)
to capture the physics of dissipation and damping in real time.
A resummation scheme is needed that effectively sums up the whole
(or partial) perturbative
series in a consistent and/or controlled
manner,
and which provides a reliable estimate for the long-time behavior.

The next section is devoted to the analytical and numerical study
of some of these schemes.

\newpage

\section{\bf Non-perturbative schemes I: Hartree approximation}

Motivated by the failure of the loop and amplitude expansions, we now proceed
to consider the equations of motion in some non-perturbative schemes. First we
study a single scalar model in the time dependent  Hartree approximation. After
this, we study an O(N) scalar theory in the large N limit. This last case
allows us to study the effect of Goldstone bosons on the time evolution of the
order parameter.

In a single scalar model described by the Lagrangian density of
eq.(\ref{Lagrangian}), the Hartree approximation is
implemented as follows. We again decompose the fields as
$\Phi^{\pm}=\phi+\psi^{\pm}$ and the
  Lagrangian density is given by eq.(\ref{plusminuslag}).
The Hartree approximation is obtained by assuming the
factorization (for both $\pm$ components)
\begin{eqnarray}
 \psi^3 ( \vec{x}, t) & \rightarrow & 3 \langle \psi^2 ( \vec{x},t) \rangle
   \psi (\vec{x},t)   \nonumber \\
 \psi^4 ( \vec{x}, t) & \rightarrow & 6 \langle \psi^2 ( \vec{x},t) \rangle
   \psi^2 (\vec{x},t)  -3  \langle \psi^2 ( \vec{x},t) \rangle^2
\end{eqnarray}
Translational invariance shows that
 $\langle (\psi^{\pm}(\vec{x},t))^2 ( \vec{x},t) \rangle$ can only be a
function of time, and because this is an equal time correlation
function, we have that:
\begin{equation}
\langle (\psi^+(\vec{x},t))^2 \rangle =
\langle (\psi^-(\vec{x},t))^2 \rangle \stackrel{\rm def}{=}
\langle \psi^2(t) \rangle
\end{equation}
The expectation value will be determined within a self-consistent
approximation.
After this factorization we find:
\begin{eqnarray}
  {\cal{L}}[\phi+\psi^+]-{\cal{L}}[\phi-\psi^-] =
     {\cal{L}}
      [\phi] &+& \left\{ \left( \frac{\delta
      {\cal{L}}}{\delta \phi}-\frac{1}{2} \lambda \phi \langle
     \psi^2 (t) \rangle \right)
     \psi^+    +  \frac{1}{2} \left( \partial_{\mu} \psi^+ \right)^2 \right.
\nonumber \\
 &-& \left. \frac{1}{2} {\cal{M}}^{2}(t) (\psi^+)^2 \right\}
      - \left\{ \left(\psi^+ \rightarrow
\psi^-\right) \right\},\label{hartreehamiltonian}
\end{eqnarray}
where
\begin{equation}
  {\cal{M}}^2 (t)  = V^{\prime \prime}(\phi) + \frac{\lambda}{2}
    \langle \psi^2 (t) \rangle = m^2 + \frac{\lambda}{2} \phi^2 (t) +
\frac{\lambda}{2}
    \langle \psi^2 (t) \rangle \label{timemass}
\end{equation}
The resulting Hartree equations  are obtained by using the tadpole method
$\langle \psi^{\pm} ( \vec{x},t) \rangle =0$ as before.
They are given by:
\begin{eqnarray}
   & &\ddot{\phi} + m^2 \phi+ \frac{\lambda}{6} \phi^3 +
 \frac{\lambda}{2} \phi \langle
    \psi^2( t ) \rangle =0 \label{hartreeequation} \\
   & &\langle \psi^2 ( t ) \rangle  =\int \frac{d^3 k}{(2 \pi)^3} \left[
    -i G^{<}_{k}(t,t) \right] =
  \int \frac{d^3 k}{( 2 \pi)^3} \frac{\left|U^+_{k} (t)
\right|^2}{2\omega_k(0)}
  \label{qfluctuation}   \\
   & &\left[ \frac{d^2}{d t^2} + \omega_{k}^2 (t) \right]
     U^+_{k} (t)  =0 ~~; ~~~~
  \omega_{k}^2 (t) =\vec{k}^2 + {\cal{M}}^2 (t) \label{modefunction}
\label{hartreeeqns}
\end{eqnarray}
The initial conditions for the mode functions are
\begin{equation}
 U^+_{k} (0) = 1 ~~ ; ~~~~~~~
 {\dot{U}}^{+}_k  (0) = -i \omega_k (0) \label{inmodefunction}
\end{equation}
It is clear that the Hartree approximation makes the
 Lagrangian density quadratic  at the expense of a
 self-consistent condition.
In the time independent case, this approximation sums
up all the ``daisy'' (or ``cactus'') diagrams and leads
to a self-consistent gap equation.

At this stage, we must point out that the Hartree
approximation is uncontrolled in this single scalar theory.
 This approximation does, however, become exact in the
$ N \rightarrow \infty $ limit of the $O(N)$ model which
we will discuss in the next section.

\subsection{\bf Renormalization}

We now analyze  the renormalization aspects within the Hartree approximation.
To study the renormalization we need to understand the divergences in this
integral
\begin{equation}
 \langle \psi^2 (t) \rangle = \int \frac{d^3 k}{(2 \pi )^3}
   \frac{ \left| U^+_k (t)  \right|^2}{2 \omega_k (0)}
\end{equation}
The divergences  will be determined from the large-k behavior of the mode
functions which obey the differential equations obtained from
(\ref{modefunction}) with
the initial conditions (\ref{inmodefunction}). By a
 WKB-type analysis (see\cite{boyveg,boyholveg} for a detailed description),
 in the $k \rightarrow \infty$ limit, we find
\begin{equation}
  \frac{ \left| U^+_k (t)  \right|^2}{2 \omega_k (0)} =
      \frac{1}{k} -\frac{1}{4 k^3} \left[ m^2 + \frac{\lambda}{2}
    \phi^2 (t)  +
   \frac{\lambda}{2} \langle \psi^{2} (t) \rangle \right] + {\cal{O}}
   ( \frac{1}{k^4} ) + \cdots  \label{renorint}
\end{equation}
Inserting these results in the integral, it is straightforward to find the
divergent terms and we find
\begin{equation}
\int \frac{d^3 k}{(2 \pi )^3}
   \frac{ \left| U^+_k (t)  \right|^2}{2 \omega_k (0)} = \frac{1}{8 \pi^2}
\Lambda^2 - \frac{1}{8 \pi^2} \ln \left(\frac{\Lambda}{\kappa} \right)   \left[
 m^2 + \frac{\lambda}{2}
 \phi^2 (t)   +
   \frac{\lambda}{2} \langle \psi^2 (t) \rangle \right] + finite
\label{divergence}
\end{equation}
where $\Lambda$ is an upper momentum cutoff and $\kappa$ a renormalization
scale.

Now, we are in  position to specify the renormalization
prescription within the Hartree approximation. In this
approximation, there are no interactions, since the
 Lagrangian density  is quadratic. The nonlinearities
are encoded in the self-consistency conditions.
Because of this, there are no counterterms with which to
cancel the divergence and the differential equation for
the mode functions must be finite. Therefore, it leads
 to the following renormalization prescription:
\begin{equation}
  m^2_B + \frac{\lambda_B}{2} \phi^2 (t) +
  \frac{\lambda_B}{2} \langle \psi^2 (t) \rangle_B =
  m^2_R + \frac{\lambda_R}{2} \phi^2 (t) +
 \frac{\lambda_R}{2} \langle \psi^2 (t) \rangle_R
\label{renormalizationprescription}
\end{equation}
where the subscripts $B,R$ refer to the bare and
renormalized quantities respectively and
 $\langle \psi^2 (t) \rangle_B $ is read
from eq.(\ref{divergence}):
\begin{equation}
\langle \psi^2 (t) \rangle_B = \frac{1}{8 \pi^2}
 \Lambda^2 -\frac{1}{8 \pi^2}
 \ln \left( \frac{\Lambda}{\kappa} \right)
\left[   m^2_R + \frac{\lambda_R}{2} \phi^2 (t) +
\frac{\lambda_R}{2} \langle \psi^2 (t) \rangle_R
\right] + \mbox{ finite }
\end{equation}
Using this renormalization prescription eq.(\ref{renormalizationprescription}),
we
obtain
\begin{eqnarray}
   && m^2_B + \frac{\lambda_B}{16 \pi^2} \Lambda^2 = m^2_R \left[ 1+
\frac{\lambda_R}{16 \pi^2} \ln \left( \frac{\Lambda}{\kappa} \right) \right] \\
  && \lambda_B = \frac{\lambda_R}{1- \frac{\lambda_R}{16 \pi^2} \ln \left(
\frac{\Lambda}{\kappa} \right)} \label{renormalizationconditions}
\end{eqnarray}
and
\begin{eqnarray}
 \langle \psi^2 (t) \rangle_R=&&\left[ \frac{1}{1- \frac{\lambda_R}{16 \pi^2}
\ln \left( \frac{\Lambda}{\kappa} \right)}  \right] \times  \nonumber \\
    &&\left\{ \int \frac{d^3 k}{(2 \pi)^3}
 \frac{\left| U^+_k (t) \right|^2}{2 \omega_k (0)}
  - \left[  \frac{1}{8 \pi^2} \Lambda^2 -\frac{1}{8 \pi^2}
 \ln \left( \frac{\Lambda}{\kappa} \right) \left(   m^2_R + \frac{\lambda_R}{2}
\phi^2 (t)    \right)  \right] \right\} \label{renormalizedfluctuation}
\end{eqnarray}
It is clear that there is no wavefunction renormalization.
This is a consequence of the approximation invoked. There
 is, in fact, no wavefunction renormalization in either
one-loop or Hartree approximation for a scalar field theory
in three spatial dimensions.

With an eye towards the  numerical analysis,
 it is more convenient to write
\begin{equation}
 \langle \psi^2 (t) \rangle_R =
\left( \langle \psi^2 (t) \rangle_R - \langle \psi^2 (0) \rangle_R \right)+
\langle \psi^2 (0) \rangle_R
\end{equation}
and perform a subtraction at time $t=0$
absorbing $\langle \psi^2 (0) \rangle_R $ into a further
{\em finite} renormalization of the mass term
($ m^2_R +\langle \psi^2 (0) \rangle_R = M^2_R $).

  The renormalized equations that  we will solve finally become
\begin{eqnarray}
   &&\ddot{\phi} + M^2_R \phi + \frac{\lambda_R}{2}
\left[ 1- \left(\frac{2}{3}
\right) \frac{1}{1- \frac{\lambda_R}{16 \pi^2}
\ln \left( \frac{\Lambda}{\kappa}\right)} \right] \phi^3 +
\frac{\lambda_R}{2} \phi \left( \langle \psi^2 (t) \rangle_R - \langle \psi^2
(0) \rangle_R \right)
 =0 \nonumber \\
{}~~~ \\
  &&\left[ \frac{d^2}{dt^2} + k^2 +M^2_R +
\frac{\lambda_R}{2} \phi^2 (t) +
\frac{\lambda_R}{2}\left( \langle \psi^2 (t) \rangle_R - \langle \psi^2 (0)
\rangle_R \right)
\right]U^+_k (t) =0 \label{rhartreeequation}
\end{eqnarray}
and
\begin{eqnarray}
\left( \langle \psi^2 (t) \rangle_R - \langle \psi^2 (0) \rangle_R \right)
 & & = \left[ \frac{1}{1- \frac{\lambda_R}{16 \pi^2} \ln \left(
\frac{\Lambda}{\kappa} \right)}\right] \times \nonumber \\
 & & \left\{ \int \frac{d^3 k}{(2 \pi)^3}
 \frac{\left| U^+_k (t) \right|^2 -1}{2 \omega_k  (0)}
   + \frac{\lambda_R}{16 \pi^2}
 \ln \left( \frac{\Lambda}{\kappa} \right) \left(    \phi^2 (t) -
  \phi^2 (0)  \right)   \right\}  \label{psisubtracted}
\end{eqnarray}
with the initial conditons for $U^+_k (t) $:
\begin{equation}
 U^{+}_{k} (0) = 1~~ ; ~~~~~~~
 {\dot{U}}^{+}_k  (0) = -i \omega_k (0) \; \; ; \; \;
\omega_k (0) = \sqrt{ k^2 +M^2_R +
\frac{\lambda_R}{2} \phi^2 (0)}
\label{initialumodefunction}
\end{equation}

It is worth noticing  that there is a weak cutoff dependence
 on the renormalized   equations of motion of the order
 parameter and the mode functions.
 This is a consequence of the well known ``triviality'' problem
of the scalar quartic interaction in four space-time dimensions.
This has the consequence  that for a fixed renormalized coupling
the cutoff must be kept fixed and finite. The presence of
the Landau pole prevents taking the limit of the ultraviolet
cutoff to infinity while keeping the renormalized coupling fixed.

 This theory is sensible only as
a low-energy cutoff effective theory. We then must be careful
that for a fixed value of $\lambda_R$, the cutoff must be such
that the theory never crosses the Landau pole.
 Thus from a numerical perspective there will always be
a cutoff sensitivity in the theory. However, for small coupling we expect the
cutoff dependence to be rather weak (this will  be confirmed numerically)
provided the cutoff is  far away from the Landau pole.

\subsection{Particle Production}

Before we engage ourselves in a numerical integration of the
above equations of motion we want to address the issue of
particle production since it is of great importance for the understanding of
dissipative processes.

In what follows, we consider  particle production due to
the time varying effective mass $ {\cal{M}}^2 (t) $ in
eq.(\ref{timemass}) of the quantum field
$ \psi $ for the single scalar model.

The Lagrangian density for the fluctuations in the Hartree approximation is
given by eq.(\ref{hartreehamiltonian}). Demanding that the order parameter
satisfies its equation of motion implies that the linear term in $\eta$ in
eq.(\ref{hartreehamiltonian}) vanishes. The resulting Lagrangian density is
\begin{equation}
{\cal{L}} [\psi]  =  \frac{1}{2} \left( \partial_{\mu} \psi \right)^2  -
     \frac{1}{2} {\cal{M}}^{2}(t) \psi^2  + \frac{\lambda}{8}\langle \psi^2
\rangle
\end{equation}
To study the issue of the particle production, it  proves
 convenient to pass to Hamiltonian density:
\begin{equation}
{\cal{H}} = \Pi_{\psi} \dot{\psi} -{\cal{L}}
    = \frac{1}{2} \Pi_{\psi}^2 + \frac{1}{2} \left( \vec{\nabla}
    \psi \right)^2 + \frac{1}{2} {\cal{M}}^2 (t) \psi^2
- \frac{\lambda}{8}\langle \psi^2 \rangle  \label{hami}
\end{equation}
Here $  \Pi_{\psi} $ is the canonical  momentum conjugate to
 $\psi$.

In a time dependent background the concept of particle is ambiguous and it
must be defined with respect to some particular state.
Let us consider the Heisenberg fields at $t=0$  written as
\begin{eqnarray}
 && \psi ( \vec{x},0) = \frac{1}{\sqrt{\Omega}} \sum_k \frac{1}{\sqrt{2
\omega_k (0)}} \left( {a}_k (0) + {a}^{\dagger}_{-k} (0) \right)  e^{i \vec{k}
\cdot \vec{x}}  \nonumber \\
 && \Pi_{\psi} ( \vec{x},0) = \frac{-i}{\sqrt{\Omega}} \sum_k
\sqrt{\frac{\omega_k (0)}{2}}
 \left( {a}_k (0) - {a}^{\dagger}_{-k} (0) \right)  e^{i \vec{k} \cdot \vec{x}}
 \label{initialexpansion}
\end{eqnarray}
with $\omega_k (0)$ as in eq.(\ref{modefunction}) and $\Omega$ the spatial
volume.
The  Hamiltonian is diagonalized at $t=0$ by these creation and
destruction operators:
\begin{equation}
 H(0) = \sum_k \omega_k (0) \left[ {a}^{\dagger}_k (0) {a}_k (0) +\frac{1}{2}
\right]
\end{equation}

Thus we define the Hartree-Fock states at $t=0$ as the vacuum annihilated by
$a_k(0)$ together with the tower of excitations obtained by applying
polynomials in $a^{\dagger}_k(0)$ to this vacuum state. The Hartree-Fock vacuum
state at $t=0$ is chosen as the reference state. As time passes, particles (as
defined with respect to this state) will be produced as a result of parametric
amplification\cite{hupa,brantra}. We should mention that our definition differs
from
that of other authors\cite{morikawa,brantra} in that we chose the state at time
$t=0$ rather than using the adiabatic modes (that diagonalize the instantaneous
Hamiltonian).

We define the number density of particles as a function of time as
\begin{equation}
{\cal{N}}(t) = \int \frac{d^3k}{(2\pi)^3}\frac{Tr\left[a^{\dagger}_k(0)a_k(0)
\rho(t)\right]}{Tr \rho(0)}=
\int \frac{d^3k}{(2\pi)^3}\frac{Tr\left[a^{\dagger}_k(t)a_k(t)
\rho(0)\right]}{Tr \rho(0)}
\label{numbdens}
\end{equation}
where by definition
\begin{equation}
a^{\dagger}_k(t) = U^{-1}(t,0)a^{\dagger}_k(0)U(t,0) \; ; \;
a_k(t) = U^{-1}(t,0)a_k(0)U(t,0) \label{heisops}
\end{equation}
are the time-evolved operators in the Heisenberg picture. In terms of these
time
evolved operators, we may write:
\begin{eqnarray}
  \psi ( \vec{x},t) & = & \frac{1}{\sqrt{\Omega}} \sum_k
\frac{1}{\sqrt{2 \omega_k (0)}} \left( {a}_k (t) + {a}^{\dagger}_{-k} (t)
\right)  e^{i \vec{k} \cdot \vec{x}}  \nonumber \\
\Pi_{\psi} ( \vec{x},t)& = & \frac{-i}{\sqrt{\Omega}} \sum_k
\sqrt{\frac{\omega_k (0)}{2}}
 \left( {a}_k (t) - {a}^{\dagger}_{-k} (t) \right)  e^{i \vec{k} \cdot \vec{x}}
 \label{timeexpansion}
\end{eqnarray}

On the other hand, we now expand the Heisenberg fields at time $t$ in the
following orthonormal basis
\begin{eqnarray}
   \psi ( \vec{x},t ) &=& \frac{1}{\sqrt{\Omega}}
\sum_{\vec{k}}\frac{1}{\sqrt{2\omega_k(0)}}
\left( \tilde{a}_k U^+_{k} (t) +
    \tilde{a}^{\dagger}_{-k} U^-_k (t) \right)
 e^{i \vec{k} \cdot \vec{x}} \nonumber \\ \label{psihar}
\Pi_{\psi} ( \vec{x},t ) &=& \frac{1}{\sqrt{\Omega}}
\sum_{\vec{k}}\frac{1}{\sqrt{2\omega_k(0)}}
 \left( \tilde{a}_k \dot{U}^{+}_{k} (t) +
    \tilde{a}^{\dagger}_{-k} \dot{U}^-_k (t) \right)
 e^{i \vec{k} \cdot \vec{x}} \label{pihar}
 \label{orthonormalexpansion}
\end{eqnarray}
where the mode functions $U^+_k(t) \; ; \; U^-_k(t)=(U^+_k(t))^*$ are the
Hartree-Fock mode functions obeying
eqs.(\ref{rhartreeequation}-\ref{initialumodefunction}) together with
the self consistency condition.

Thus $\tilde{a}_k$, $\tilde{a}^{\dagger}_k$ are the   annihilation and creation
operators of Hartree-Fock states, and the Heisenberg field  $\psi (\vec{x},t) $
is a solution of the Heisenberg equations of motion in the Hartree
approximation.  Therefore the  $\tilde{a}_k^{\dagger}$ and $\tilde{a}_k$ do not
depend  on time and are identical to $a_k^{\dagger} (0)$ and $a_k (0)$
respectively ( we can check this  by evaluating the  expansion in
eq.(\ref{orthonormalexpansion}) at $t=0$ together with the  initial conditions
on $U^+_k (t) $ in eq.(\ref{inmodefunction})). Using the Wronskian properties
of the function $U^+_k (t)$, we see that the  $\tilde{a}_k^{\dagger}$ and
$\tilde{a}_k$ satisfy the usual  canonical commutation relations. The reason
for  the choice of the vacuum state at $t=0$ now becomes clear; this is the
initial time at  which the boundary conditions on the modes are determined.The
mode functions  $U^+_k(t)\; ; \;U^-_k(t) $ are then identified with positive
and negative  frequency modes at the initial time.

By comparing the expansion in eq.(\ref{initialexpansion})
 evaluated at time $t$ with that in
eq.(\ref{orthonormalexpansion}), we find that the creation and
annihilation operator at time $t$ can be related to those at
 time $t=0$ via a Bogoliubov transformatiom:
\begin{equation}
   a_k (t) = {\cal{F}}_{+,k} (t) {a}_k (0) +{\cal{F}}_{-,k}
(t) {a}^{\dagger}_{-k} (0)
\end{equation}
The $ {\cal{F}}_{\pm} (t)$ can be read off in terms of the
mode functions  $U^+_k (t)$
\begin{eqnarray}
 \left| {\cal{F}}_{+,k} (t) \right|^2 &=&
\frac{1}{4} \left| U^+_k(t) \right|^2 \left[ 1+
  \frac{ \left| \dot{U}^+_k (t) \right|^2}{\omega^2_k  (0)
\left| U^+_k(t) \right|^2} \right] +\frac{1}{2} \nonumber \\
   \left| {\cal{F}}_{+,k}(t) \right|^2    &-& \left| {\cal{F}}_{-,k} (t)
\right|^2
  =1
   \label{bogoluibov}
\end{eqnarray}

At any time $t$ the expectation value of the number operator for
the quanta of $\psi$ in each k-mode is given by
\begin{equation}
{\cal{N}}_k(t) = \frac{Tr\left[a^{\dagger}_k(t)a_k(t)\rho(0)\right]}{Tr
\rho(0)}
\end{equation}

After some algebra, we find
\begin{equation}
  {\cal{N}}_k(t) = \left( 2  \left| {\cal{F}}_{+,k}(t) \right|^2 -1
 \right)  {\cal{N}}_k (0) + \left( \left| {\cal{F}}_{+,k}(t) \right|^2
-1 \right) \label{numberoperator}
\end{equation}
This result exhibits the contributions from ``spontaneous''
 ( proportional to the initial particle occupations) and
``induced'' ( independent of it) particle production. Since
we are analyzing the zero temperature case with ${\cal{N}}_k(0)=0$ only
the induced contribution results.

Before passing on to the numerical analysis it is important to point out that
the Hartree approximation is {\em energy conserving}. The bare energy density
is
\begin{equation}
{\cal{E}}=\frac{1}{2}{\dot{\phi}}^2(t)+V(\phi(t))+\frac{Tr{\cal{H}}\rho(0)}{Tr\rho(0)} \label{energy}
\end{equation}
with $V(\phi)$ the classical potential and ${\cal{H}}$ is given by
eq.(\ref{hami}) with $\Pi_{\psi} \; , \; \psi$ expanded in terms of the Hartree
mode
functions and creation and annihilation operator as in eqs.(\ref{psihar},
\ref{pihar}). Using the equations of motion for $\phi(t)$ and the mode
functions, and after some lengthy but straightforward algebra one finds
$\dot{\cal{E}} =0$.

\subsection{Numerical Analysis}

\subsubsection{Unbroken Symmetry Case}

In order to perform a numerical analysis it is necessary
to introduce dimensionless quantities and it becomes convenient
to chose the renormalization point $\kappa = M_R$. Thus we define
\begin{eqnarray}
\eta(t) & = & \phi(t) \sqrt{\frac{\lambda_R}{2 M^2_R}}
\; ; \;
q= \frac{k}{M_R} \; ; \; \tau =  M_R t \; ; \; g= \frac{\lambda_R}{8\pi^2}
\nonumber \\
\Sigma(t)
  & = &  \frac{4 \pi^2}{M^2_R}
\left( \langle \psi^2 (t) \rangle_R - \langle \psi^2 (0)
 \rangle_R \right) \label{dimensionless2}
\end{eqnarray}
and finally, the equations of motion become
\begin{eqnarray}
   &&\frac{d^2}{d\tau^2}{\eta} + \eta +
\left[ 1- \left(\frac{2}{3}
\right) \frac{1}{1- \frac{g}{2}
\ln \left( \frac{\Lambda}{M_R}\right)} \right] \eta^3 +
g \eta \Sigma(\tau)
 =0 \nonumber \\
  &&\left[ \frac{d^2}{d\tau^2} + q^2 +1 +
 \eta^2 (\tau) +
g \Sigma(\tau)
\right]U^+_q (\tau) =0 \label{dimrhartreeequation1} \\
  && U^+_q (0)=1 \; ; \; \frac{d}{d \tau}U^+_q (0) = -i
\sqrt{q^2 +1 + \eta^2 (0)} \label{dimbounharcon}
\end{eqnarray}
\begin{eqnarray}
\Sigma(\tau) =  &&
\left[ \frac{1}{1- \frac{g}{2} \ln \left( \frac{\Lambda}{M_R} \right)}\right]
\times \nonumber \\
                       &&\left\{ \int^{\Lambda / M_R}_0 q^2 dq
 \frac{\left| U^+_q (\tau) \right|^2 -1}
{\sqrt{ q^2 +1 + \eta^2 (0)}}
   + \frac{1}{2}
 \ln \left( \frac{\Lambda}{M_R} \right) \left( \eta^2 (\tau)
 -   \eta^2 (0)  \right)   \right\} \label{sigmatau}
\end{eqnarray}

In terms of the dimensionless quantities we obtain the number of
particles within a {\em correlation volume} $N(\tau)={\cal{N}}(t)/M^3_R$
\begin{equation}
N(\tau)=\frac{1}{8\pi^2}\int^{\Lambda / M_R}_0 q^2 dq
\left\{
  \left| U^+_q(\tau) \right|^2+
  \frac{ \left| \dot{U}^+_q (\tau) \right|^2}{\sqrt{ q^2 +1 + \eta^2 (0)}}
  -2 \right\} \label{particleproduction}
\end{equation}

Figures (5.a,b,c) show $\eta(\tau)$, $\Sigma(\tau)$ and  $N(\tau)$ in the
Hartree approximation, for $g=0.1$, $\eta(0)=1.0$ and $\Lambda / M_R =100$; we
did not detect an appreciable cutoff dependence by varying the cutoff between
50 and 200. Clearly there is no appreciable damping in $\eta(\tau)$. In fact it
can be seen that the period of the oscillation is very close to $2\pi$, which
is the period of the classical solution of the linear theory. This is
understood because the coefficient of the cubic term is very small and $g
\Sigma(\tau)$ is negligible. Particle production is also negligible. This
situation should be contrasted with that shown in figures (6.a-c) and (7.a-c)
in which there is  dissipation and damping in the evolution of $\eta(\tau)$ for
$\eta(0)=4,\ 5$ respectively and the same values for $g$ and the cutoff. There
are several noteworthy features that can be deduced from these figures. First
the fluctuations become very large, such that $g\Sigma(\tau)$ becomes
${\cal{O}}(1)$. Second, figures (6.a) and (7.a) clearly show that initially,
channels are open and energy is transferred away from the $q=0$ mode of the
field. Eventually
however, these channels shut off, and the dynamics of the expectation
value settles
into an oscillatory motion. The time scale for the shutting off of the
dissipative behavior decreases as $\eta(0)$ increases; it is about 25 for
$\eta(0)=4$, and about 18 for $\eta(0)=5$. This time scale is correlated with
the time scale in which particles are produced by parametric
amplification and the quantum fluctuations begin to plateau (figures (6.b,c)
and (7.b,c)). Clearly the dissipative mechanism which damps the motion of the
expectation value is particle
production. Furthermore, the long time dynamics for the expectation value
{\em does
not} correspond to exponential damping. In fact, we did not find any
appreciable
damping for $\tau \geq 70$ in these cases. It is illuminating to compare this
situation with that of a smaller coupling depicted in figures (8.a-c) for
$\eta(0)=5$ $g=0.05$. Clearly the time scale for damping is much larger and
there
is still appreciable damping at $\tau \approx 100$. Notice also that
$g\Sigma(\tau) \approx 2$ and that particles are being produced even at long
times and this again correlates with the evidence that the field expectation
value shows
damped motion at long times, clearly showing that the numerical analysis has
not reached the asymptotic regime.

These numerical results provide some interesting observations. First we
see  that
$\eta^2(\tau)+g\Sigma(\tau)$ has asymptotic oscillatory behavior in $\tau$.
 In general
this would imply unstable bands of wavevectors for the
solutions of equation (\ref{dimrhartreeequation1}).
 However, if these unstable bands are present, the mode functions would grow
and
and $\Sigma(\tau)$ would grow as a result since it is an integral over all
wavevectors.
The fact that this does not happen implies that there are no
unstable bands.

We have been able to find analytically an exact oscillatory solution of the
equation (\ref{dimrhartreeequation1}) which is given by
\begin{eqnarray}
1+\eta^2(\tau)+g\Sigma(\tau) & = & -e_3-2 \; {\cal{P}}(\tau+\omega')
\label{massex} \\
\eta(\tau)                   & = & A \; {\rm sn}(\sqrt{e_1-e_3}\tau)
\end{eqnarray}
where ${\cal{P}}$ is the Weierstrass function which is  double periodic
 with periods
$\omega \; ; \; \omega'$\cite{abram}. The parameters in this solution are
functions of the coupling. We found that, asymptotically, the numerical
solution of equations (\ref{dimrhartreeequation1}, \ref{dimbounharcon}) could
be fit quite precisely to the above analytic solution for precise values of the
parameters.  These values encode the information about the initial conditions
and
the coupling constant.
For this potential given by (\ref{massex}) we can find the exact solutions for
the mode functions in terms of Weierstrass $\sigma$ functions\cite{ince}. In
this case
there are no unstable modes for real wavector $q$.
This is an important result
because it provides an analytic understanding for the lack of forbidden
bands.

Although this exact result does not illuminate the physical reason why there
are no unstable bands, we conjecture that this is a consequence of the
conservation
of energy in the Hartree approximation.

Numerically we find no evidence for exponential relaxation asymptotically, of
course,
we cannot rule out the possibility of power law asymptotic decay for $\eta$
with
an exceedingly small power.

The fundamental question to be raised at this point is: what is the
origin of the damping in the evolution of the field expectation value? Clearly
this is a collisionless process as collisions are not taken into
account in the Hartree approximation (although the one-loop
 diagram that enters in the two particle collision amplitude
with  the two-particle cut is contained in the Hartree approximation and is
responsible for thresholds to particle production).
 The physical mechanism is very similar to that
of Landau damping in the collisionless Vlasov equation for plasmas\cite{landau}
and also found in the study of strong electric fields in
reference\cite{klug}. In the case under consideration, energy
 is transfered from the expectation value
to the quantum fluctuations which back-react on the evolution of
the field expectation value but out of phase.  This phase difference between
the oscillations
of $\eta^2(\tau)$ and those of $\Sigma(\tau)$ can be clearly seen to be
$\pi$ in figures (6.a,b), (7.a,b) since the maxima of $\eta^2(\tau)$ occur at
the same times as
the minima of $\Sigma(\tau)$ and viceversa.

This is an important point learned
from our analysis and that is not {\em a priori} taken into account in the
usual arguments for dissipation via collisions. The process of
thermalization, however, will necessarily involve collisions and
cannot be studied within the schemes addressed in this paper.

\subsubsection{Broken Symmetry}

The broken symmetry case is obtained by writing $M^2_R = -\mu_R^2 < 0$ and
using
the scale $\mu_R$ instead of $M_R$ to define the dimensionless
quantities as in eq.(\ref{dimensionless2}) and the renormalization scale.
The equations of motion in this case become
\begin{eqnarray}
   &&\frac{d^2}{d\tau^2}{\eta} - \eta +
\left[ 1- \left(\frac{2}{3}
\right) \frac{1}{1- \frac{g}{2}
\ln \left( \frac{\Lambda}{\mu_R}\right)} \right] \eta^3 +
g \eta \Sigma(\tau)
 =0 \nonumber \\
  &&\left[ \frac{d^2}{d\tau^2} + q^2 -1 +
 \eta^2 (\tau) +
g \Sigma(\tau)
\right]U^+_q (\tau) =0 \label{dimbrorhartreeequation} \\
  && U^+_q (0)=1 \; ; \; \frac{d}{d \tau}U^+_q (0) = -i
\sqrt{q^2 +1 + \eta^2 (0)} \label{dimbouncon}
\end{eqnarray}
with $\Sigma(\tau)$ given in eq.(\ref{sigmatau}) but with $M_R$
replaced by $\mu_R$. The broken symmetry case is more subtle
because of the possibility of unstable modes for  initial
conditions in which $\eta(0) <<1$. We have kept the boundary conditions
of eq.(\ref{dimbouncon}) the same as in eq.(\ref{dimbounharcon}). This
corresponds to preparing an initial state as a Gaussian state
centered at $\eta(0)$ with real and positive covariance (width)
and letting it evolve for $t>0$ in the
broken symmetry potential\cite{boyveg,dcc}.
The number of particles produced within a correlation volume (now $\mu_R^{-3}$)
is
given by eq.(\ref{particleproduction}) with $M_R$ replaced by $\mu_R$.

Figures (9.a-c) depict the dynamics for a broken symmetry case
in which $\eta(0)=10^{-5}$ i.e. very close to the top of the potential
hill. Notice that as the field expectation value rolls down the hill the
unstable modes
make the fluctuation grow dramatically until about $\tau \approx 50$ at
which point $g\Sigma(\tau) \approx 1$. At this point, the unstable
growth of fluctuations shuts off and the field begins damped oscillatory
motion around a mean value of about $\approx 1.2$.
This point is a minimum of the effective action.
 The damping of these
oscillations is very similar to the damping around the origin in the
unbroken case. Most of the particle production and the largest quantum
fluctuations occur when the field expectation value
is  rolling down the  region for which there are unstable
frequencies for the mode functions
 (see eq.(\ref{dimbrorhartreeequation})).
This behavior is similar to that found previously by some of these
authors\cite{boyveg}.

\section{\bf Non-perturbative schemes II: Large N limit in the $ O(N)$ Model}

Although the Hartree approximation offers a non-perturbative resummation of
select terms, it is not  a consistent approximation because there is no
{\it a priori} small parameter that defines the approximation. Furthermore
we want to study the effects of dissipation by Goldstone bosons in a non
perturbative but controlled expansion.

In this section, we consider the $O(N)$ model in the large N limit.
The large N limit has been used in studies of non-equilibrium
dynamics\cite{mottola1,mottola2,mottola3,dcc} and it provides a very powerful
tool for studying non-equilibrium dynamics non-perturbatively in a consistent
manner.
The Lagrangian density is
the following:
\begin{eqnarray}
 {\cal{L}} &=& \frac{1}{2} \partial_{\mu} \vec{\phi} \cdot \partial^{\mu}
    \vec{\phi} - V(\sigma, \vec{\pi} ) \nonumber  \\
   V(\sigma, \vec{\pi} ) &=& \frac{1}{2} m^2 \vec{\phi} \cdot \vec{\phi} +
   \frac{ \lambda}{8N} (  \vec{\phi} \cdot \vec{\phi} )^2
\label{onLagrangian}
\end{eqnarray}
for $\lambda$ fixed in the large $N$ limit. Here $\vec{\phi}$ is an $O(N)$
vector,  $\vec{\phi} = (\sigma, \vec{\pi} )$ and $\vec{\pi}$ represents the
$N-1$ pions. In what follows, we will consider two different cases of the
potential $  V(\sigma, \vec{\pi} ) $, with ($m^2 < 0$) or without ($m^2 > 0$)
symmetry breaking.

Let us define the fluctuation field $\chi( \vec{x},t )$ as
\begin{equation}
   \sigma  (\vec{x},t ) = \sigma_0(t)+ \chi ( \vec{x},t)
\end{equation}
Expanding the Lagrangian density in eq.(\ref{onLagrangian}) in
terms of fluctuations $\chi(\vec{x},t)$,we obtain

\begin{eqnarray}
&& {\cal{L}} [ \sigma_0 +\chi^+, \vec{\pi}^+ ]
 -  {\cal{L}} [\sigma_0+\chi^-, \vec{\pi}^- ] =
 \left\{ {\cal{L}} [ \sigma_0,\vec{\pi}^+ ]
     +  \frac{\delta {\cal{L}} }
{\delta \sigma_0} \chi^+
   + \frac{1}{2} \left( \partial_{\mu} \chi^+ \right)^2
 \right.
    \nonumber \\
&& +
\frac{1}{2} \left( \partial_{\mu} \vec{\pi}^+ \right)^2
  \left. -\left( \frac{1}{2!} V^{\prime \prime}
 (\sigma_0,\vec{\pi}^+ ) \chi^{+2} + \frac{1}{3!} V^{[3]}
(\sigma_0,\vec{\pi}^+ ) (\chi^+)^3 + \frac{1}{4!} V^{[4]}
 (\sigma_0, \vec{\pi}^+ ) (\chi^+)^4 \right) \right\} \nonumber \\
&&- \left\{ \left( \chi^+ \rightarrow \chi^- \right), \left( \vec{\pi}^+
\rightarrow \vec{\pi}^-  \right) \right\}
\end{eqnarray}
The tadpole condition $\langle \chi^{\pm}(\vec{x},t) \rangle =0$ will lead to
the equations of motion as previously discussed.
We now introduce a Hartree factorization. In the presence of a
non-zero expectation value the Hartree factorization is subtle in the
case of continuous symmetries. A naive Hartree factorization violates the
Ward identities related to Goldstone's theorem. This shortcoming is overcome
if the Hartree factorization is implemented in leading order in the large N
expansion\cite{dcc}. We
will make a series of assumptions that seem to be reasonable,
but that can only be justified {\it a posteriori}
when we recognize that with these assumptions, we obtain the equations of
motion that fulfill the Ward identities.
These assumtions are: i) no cross correlations between the pions and sigma
field and ii) that the two point  correlation functions of the pion field are
diagonal in the $ O(N-1)$ space given by the remaining unbroken symmetry group.
Based upon  these  assumptions we are led to the following Hartree
factorization of the non-linear terms in the Lagrangian density (again for both
$\pm$ components):
\begin{eqnarray}
   \chi^4  & \rightarrow & 6 \langle \chi^2 \rangle \chi^2 + constant \\
   \chi^3  & \rightarrow & 3 \langle \chi^2 \rangle \chi \\
   \left( \vec{\pi} \cdot \vec{\pi} \right)^2 & \rightarrow & \left( 2+
    \frac{4}{N-1} \right) \langle \vec{\pi}^2 \rangle \vec{\pi}^2 + constant \\
   \vec{\pi}^2 \chi^2 & \rightarrow & \langle \vec{\pi}^2  \rangle \chi^2
  +\vec{\pi}^2 \langle \chi^2 \rangle \\
   \vec{\pi}^2 \chi & \rightarrow & \langle \vec{\pi}^2  \rangle \chi
\end{eqnarray}
where by ``constant'' we mean the operator independent expectation values of
the composite operators which will not enter into the time evolution equation
of the order parameter.

In this approximation, the resulting Lagrangian density  is quadratic, with a
linear term in $\chi$ :
\begin{eqnarray}
  {\cal{L}} [\sigma_0+\chi^+, \vec{\pi}^+ ] &-&  {\cal{L}}
[\sigma_0+\chi^-, \vec{\pi}^- ] =
 \left\{ \frac{1}{2} \left( \partial_{\mu} \chi^+ \right)^2 +\frac{1}{2} \left(
\partial_{\mu} \vec{\pi}^+ \right)^2 - \chi^+ V^{\prime} (t) \right. \nonumber
\\
  &-& \frac{1}{2}{\cal{M}}^2_{\chi} (t) (\chi^+)^2
  -\left.\frac{1}{2}{\cal{M}}^2_{\vec{\pi}} (t) (\vec{\pi}^+)^2 \right\}
    - \left\{ \left( \chi^+ \rightarrow \chi^- \right),\left( \vec{\pi}^+
\rightarrow \vec{\pi}^-  \right)\right\} \nonumber \\
\end{eqnarray}
where $V^{\prime}$ is the derivative of the Hartree potential defined below.
To obtain a large N limit, we define
 \begin{eqnarray}
 \langle
\vec{\pi}^2 \rangle &=& N \langle
\psi^2 \rangle  \label{largeN} \\
\sigma_0(t)         &=& \phi(t) \sqrt{N} \label{filargeN}
\end{eqnarray}
with
\begin{equation}
  \langle
\psi^2 \rangle \approx {\cal{O}} (1) \; ; \;  \langle
\chi^2 \rangle \approx {\cal{O}} (1) \; ; \;  \phi \approx  {\cal{O}} (1).
\label{order1}
\end{equation}
 We will approximate further by neglecting the $ {\cal{O}} (\frac{1}{N})$
 terms in the formal large $N$ limit. We now obtain
\begin{eqnarray}
   V^{'} (\phi(t),t)         &=&{\sqrt{N}}
  \phi (t) \left[ m^2 +\frac{\lambda}{2} \phi^2 (t) + \frac{\lambda}{2} \langle
\psi^2 (t) \rangle \right]
\label{vprime} \\
  {\cal{M}}^2_{\vec{\pi}}(t) &=& m^2 + \frac{\lambda}{2} \phi^2 (t) +
  \frac{\lambda}{2}\langle \psi^2 (t) \rangle  \\
  {\cal{M}}^2_{\chi}(t)      &=& m^2 + \frac{3 \lambda}{2} \phi^2 (t) +
  \frac{\lambda}{2} \langle \psi^2 (t) \rangle
\end{eqnarray}
Using the tadpole method, we obtain the following set of equations:
\begin{eqnarray}
&& \ddot{\phi}(t) +  \phi (t) \left[ m^2 +\frac{\lambda}{2} \phi^2 (t) +
\frac{\lambda}{2} \langle \psi^2 (t) \rangle  \right]
   =0 \; ; \;  \langle \psi^2 ( t ) \rangle =
   \int \frac{d^3 k}{( 2 \pi)^3} \frac{\left| U^{+}_k (t) \right|^2
}{2\omega_{\vec{\pi} k}^2 (0)}
    \label{pionmodefunction}  \\
&& \left[ \frac{d^2}{d t^2} + \omega_{\vec{\pi} k}^2 (t) \right]
     U^{+}_{k} (t)  =0 \; ; \;
  \omega_{\vec{\pi} k}^2 (t) = \vec{k}^2 + {\cal{M}}_{\vec{\pi}}^2 (t)
\end{eqnarray}
It is clear from the above equations that the Ward identities of Goldstone's
theorem
are indeed fulfilled. Because
$V^{'}(\phi (t),t)=\sqrt{N}\phi(t){\cal{M}}_{\vec{\pi}}^2(t)$,
whenever $V'(\phi(t),t)$ vanishes for $\phi \neq 0$ then
 ${\cal{M}}_{\vec{\pi}}=0$ and the
``pions'' are the Goldstone bosons.
The initial conditions for the mode functions $U^+_k (t) $ are
\begin{equation}
 U^{+}_{k} (0) = 1 ~~ ; ~~~~~~~
 {\dot{U}}^+_k  (0) = -i {\omega_{\vec{\pi} k} (0)}
\label{oninmodefunctions}
\end{equation}
Since in this approximation, the dynamics for the $\vec{\pi}$ and $\chi$
fields decouples, and the dynamics of $\chi$ does not influence that of
$\phi$ or the mode functions and $\langle \psi^2 \rangle$
 we will only concentrate on the solution for the $\vec{\pi}$ fields.

\subsection{Renormalization}

The renormalization procedure is exactly the same as that for the Hartree
case in the previous section
(see eqs. (\ref{renorint}-\ref{renormalizedfluctuation}).
We carry out the same renormalization prescription and subtraction at $t=0$ as
in the last section.
Thus we find
the following equations of motion

\begin{eqnarray}
   &&\ddot{\phi} + M^2_R \phi + \frac{\lambda_R}{2}  \phi^3
+\frac{\lambda_R}{2} \phi \left(  \langle \psi^2 (t) \rangle_R - \langle \psi^2
(0) \rangle_R  \right) =0  \nonumber\\ \label{eqofmotlarN}
   &&\left[ \frac{d^2}{dt^2} + k^2 +M^2_R + \frac{\lambda_R}{2} \phi^2 (t) +
\frac{\lambda_R}{2} \left( \langle \psi^2 (t) \rangle_R - \langle \psi^2 (0)
\rangle_R \right)  \right] U^+_{k} (t) =0 \label{onrehartreeequation}
\end{eqnarray}
with the initial conditions given by eq.(\ref{oninmodefunctions})
and
with the subtracted expectation value given by eq.(\ref{psisubtracted}).

In contrast with the Hartree equations in the previous section,
the cutoff dependence in the term proportional to $\phi^3$
in eq.(\ref{eqofmotlarN})
has disappeared. This is a consequence of the large N limit and the Ward
identities, which are now obvious at the level of the renormalized equations of
motion. There is still a very weak cutoff dependence in
eq.(\ref{psisubtracted}) because of the triviality issue which is not
relieved in the large N limit, but again, this theory only makes sense as a
low-energy cutoff theory.

\subsection{Numerical Analysis}

\subsubsection{Unbroken symmetry}
To solve the evolution equations in
eqs.(\ref{psisubtracted},\ref{onrehartreeequation})  numerically, we
now introduce dimensionless quantities as in eq.(\ref{dimensionless2})
obtaining the following dimensionless equations:
\begin{eqnarray}
   &&\frac{d^2}{d\tau^2}{\eta} + \eta +   \eta^3 +
g \eta \Sigma(\tau)
 =0 \nonumber \\
  &&\left[ \frac{d^2}{d\tau^2} + q^2 +1 +
 \eta^2 (\tau) +
g \Sigma(\tau)
\right]U^+_q (\tau) =0 \label{dimrhartreeequation} \\
  && U^+_q (0)=1 \; ; \; \frac{d}{d \tau}U^+_q (0) = -i
\sqrt{q^2 +1 + \eta^2 (0)} \label{dimbocon}
\end{eqnarray}
\begin{eqnarray}
\Sigma(\tau) =  &&
\left[ \frac{1}{1- \frac{g}{2} \ln \left( \frac{\Lambda}{M_R} \right)}\right]
\times \nonumber \\
    &&\left\{ \int^{\Lambda / M_R}_0 q^2 dq
 \frac{\left[\left| U^+_q (\tau) \right|^2 -1\right]}
{\sqrt{ q^2 +1 + \eta^2 (0)}}
   + \frac{1}{2}
 \ln \left( \frac{\Lambda}{M_R} \right) \left( \eta^2 (\tau) -
 \eta^2 (0)  \right)   \right\} \label{sigmun}
\end{eqnarray}

For particle production in the $O(N)$ model, the final
expression of the  expectation value of the number operator
 for each pion field in terms of dimensionless
 quantities is the same as in eq.(\ref{particleproduction}),
but the mode function $U^+_q (t)$ obeys the differential
equation in eq.(\ref{dimrhartreeequation1}) together with the
self consistent condition.

Figures (10.a-c), (11.a-c) and (12.a-c) show $\eta(\tau)\; ; \; \Sigma(\tau)\;
; \;
N(\tau)$ for $\eta(0)=1;2$  $g=0.1;0.3$ and $\Lambda / M_R=100$
(although again we did not find cutoff sensitivity). The dynamics is very
similar to that of the single scalar field in the Hartree approximation,
which is not
surprising, since the equations are very similar (save for the coefficient
of the cubic term in the equation for the field expectation value).  Thus the
analysis presented
previously for the Hartree approximation remains valid in this case.

\subsubsection{Broken symmetry}
The broken symmetry case corresponds to choosing $M^2_R = -\mu^2_R <0$. As in
the
case of the Hartree approximation, we now choose $\mu_R$ as the scale to
define dimensionless variables and renormalization scale. The equations of
motion
for the field expectation value and the mode functions now become
\begin{eqnarray}
   &&\frac{d^2}{d\tau^2}{\eta} - \eta +   \eta^3 +
g \eta \Sigma(\tau)
 =0 \nonumber \\
  &&\left[ \frac{d^2}{d\tau^2} + q^2 -1 +
 \eta^2 (\tau) +
g \Sigma(\tau)
\right]U^+_q (\tau) =0 \label{largenmod}
\end{eqnarray}
with $\Sigma(\tau)$ given by eq.(\ref{sigmun}).
As in the  Hartree case, there is a subtlety with the boundary conditions for
the mode functions because the presence of  the instabilities at $\tau=0$ for
the band of wavevectors $0 \leq q^2 <1$, for $\eta^2(0)<1$. Following the
discussion in the
Hartree case (broken symmetry) we chose the initial conditions for the mode
functions as
\begin{equation}
 U^+_q (0)=1 \; ; \; \frac{d}{d \tau}U^+_q (0) = -i
\sqrt{q^2 +1 + \eta^2 (0)}
\end{equation}
These boundary conditions correspond to preparing a gaussian state centered
at $\eta(0)$ at $\tau=0$ and letting this initial state evolve in time in the
``broken symmetry potential''\cite{boyveg}.

Figures (13.a-c) show the dynamics of $\eta(\tau)$, $\Sigma(\tau)$
 and $N(\tau)$  for $\eta(0)=0.5$ $g=0.1$
and cutoff $\Lambda / \mu_R=100$. Strong damping behavior is evident, and the
time scale of damping is correlated with the time scale for growth of
$\Sigma(\tau)$
and $N(\tau)$. The asymptotic value of $\eta(\tau)$ and $\Sigma(\tau)$,
$\eta(\infty)$ and $\Sigma(\infty)$ respectively satisfy
\begin{equation}
-1 +  \eta^2 (\infty) +
g \Sigma(\infty)=0 \label{goldsto}
\end{equation}
as we confirmed numerically. Thus the mode functions are ``massless''
describing
Goldstone bosons. Notice that this value also corresponds to $V'(\phi)=0$
in eq.(\ref{vprime}). An equilibrium self-consistent solution of the
equations of motion for the field expectation value and the fluctuations is
reached for
$\tau= \infty$.

Figure (13.d) shows the number of particles produced per
correlation volume as a function of (dimensionless) wavevector $q$
 at $\tau=200$.
We see  that it is strongly peaked at $q=0$ clearly showing that the particle
production mechanism is most efficient for long-wavelength Goldstone bosons.
Figures (13.e-g) show snapshots of the number of particles as a function of
wavevector for $\tau=13,25,50$ respectively, notice the scale. Clearly at
longer times, the contributions from $q \neq 0$ becomes smaller.
Figures (14.a-c) show the evolution of $\eta(\tau)$, $\Sigma(\tau)$ and
$N(\tau)$
for $\eta(0)=10^{-4}$, $g=10^{-7}$. The initial value of $\eta$ is very
close to the ``top'' of the potential. Due to the small coupling and the
small initial value of $\eta(0)$, the unstable modes (those for which
$q^2 < 1$) grow for a long time making the fluctuations very large. However
the fluctuation term $\Sigma(\tau)$ is multiplied by a very small coupling
and it has to grow for a long time to overcome the instabilities. During
this time the field expectation value rolls down the potential hill, following
a trajectory
very close to the classical one.
  The
classical turning point of the trajectory beginning very near  the top
of the potential hill, is close to $\eta_{tp}=\sqrt{2}$.  Notice that
$\eta(\tau)$ exhibits a turning point (maximum)
 at $\eta \approx 0.45$. Thus the turning point of the effective
evolution equations is much
closer to the origin. This phenomenon shows that the effective (non-local)
 potential is shallower than the classical potential, with the minimum moving
closer to the origin as a function of time.

If the energy for the field expectation value was absolutely conserved,
the expectation value of the scalar field would bounce back to the initial
point and oscillate between the
two classical turning points. However, because
the fluctuations are growing and energy is transferred to them from the
$q=0$ mode, $\eta$ is slowed down as it bounces back, and tends to settle at a
value very close to the origin (asymptotically about 0.015). Figure (14.b)
shows that the fluctuations grow initially and stabilize at a value for which
$g\Sigma(\infty) \approx 1$. The period of explosive growth of the fluctuations
is correlated with the strong oscillations at the maximum of $\eta(\tau)$.
This is the time when the fluctuations begin to effectively absorb the energy
transferred by the field expectation value and when the damping mechanism
begins to work.
Again the asymptotic solution is such that $-1 +  \eta^2 (\infty) +
g \Sigma(\infty)=0 $, and the particles produced are indeed Goldstone bosons
but the value of the scalar field in the broken symmetry  minimum is very
small
(classically it would be $\eta_{min}=1$, yet dynamically, the field
settles at a value $\eta(\infty) \approx 0.015$!! for $g=10^{-7}$).
Figure (14.c) shows copious particle production, and the asymptotic final state
is a highly excited state with a large number (${\cal{O}}(10^5)$)
 of Goldstone bosons per correlation volume.

The conclusion that we reach from the numerical analysis is that Goldstone
bosons are {\it extremely effective} for dissipation and damping. Most of the
initial potential energy of the field is converted into particles (Goldstone
bosons) and the field expectation value comes to rest at long times at a
position very close
to the origin.

Notice that the difference with the situation depicted in figures (13.a-c)
is in the initial conditions and the strength of the coupling. In the case
of stronger coupling, the fluctuations grow only for a short time because
$g\Sigma(\tau)$ becomes ${\cal{O}}(1)$ in short time, dissipation begins to
act rather rapidly and the expectation value rolls down only for a short span
and
comes to rest at a minimum of the effective action,
having transferred all of its potential energy difference to produce Goldstone
bosons.

Figures (15.a-c) show a very dramatic picture. In this case
$\eta(0)=10^{-4}$,
$g = 10^{-12}$. Now the the field begins very close to the top of the
potential hill. This initial condition corresponds to a ``slow-roll''
scenario. The fluctuations must grow for a long time before $g\Sigma(\tau)$
becomes ${\cal{O}}(1)$, during which the field expectation value evolves
classically
reaching the classical turning point at $\eta_{tp} = \sqrt{2}$,
and then bouncing back. But
by the time it gets near the origin again, the fluctuations have grown
dramatically absorbing most of the energy of the field expectation value and
completely
damping its motion. In this case {\em almost all} the initial
potential energy has been converted into particles.
This is a remarkable result. The conclusion of this analysis is that the strong
dissipation by Goldstone bosons dramatically changes the dynamics of the phase
transition. For slow-roll initial conditions the scalar field relaxes to a
final value which is very close to the origin. This is the minimum of the
effective
action, rather than the minimum of the tree-level effective potential. Thus
dissipative effects by Goldstone bosons introduce a very strong dynamical
correction
of the effective action leading to a very shallow
effective potential (the effective action for constant field configuration).
The condition for this situation to happen is that the period of the classical
trajectory is of the same order of magnitude as the time scale of growth for
the fluctuations.

The weak coupling estimate for this dynamical non-equilibrium time scale
is $\tau_s \approx \ln(1/g)/2$, which is
obtained by requiring that $g\Sigma(\tau) \approx 1$. For weak coupling the
mode functions grow as $U^+_q(\tau) \approx e^{\tau}$ for $q^2 <1$, and
$\Sigma(\tau) \approx e^{2\tau}$. The numerical analysis confirms this time
scale for weak coupling. For weakly coupled theories, this non-equilibrium
 time
scale is much larger than the static correlation length (in units in which
$c=1$) and the only relevant time scale for the dynamics.

Clearly the outcome of the non-equilibrium evolution will depend on the initial
conditions of the field expectation value.

Our results pose a fundamental question: how is it possible to reconcile
damping and dissipative behavior, as found in this work with time reversal
invariance?

In fact we see no contradiction for the following reason: the dynamics is
completely determined by the set of equations of motion for the field
expectation value and
the mode functions for the fluctuations described above. These equations are
solved by providing non equilibrium initial conditions on the field expectation
value, its derivative and the
mode functions and their derivatives at the initial time $t=0$. The problem is
then evolved in time by solving the coupled {\em second order} differential
equations. We emphasize the fact that the  equations are second order in time
because these are time reversal invariant. Now consider evolving this set of
equations up to a positive time $t_0$, at which we stop the integration and
find the value of the field expectation value, its derivative, the value of
{\em all} the
mode functions and their derivatives. Because this is a system of
differential equations which is second order in time, we can take these values
at $t_0$ as  initial
conditions at this time and evolve {\em backwards} in time reaching the initial
values  at $t=0$. Notice that doing this involves beginning at a time $t_0$ in
an excited state with (generally) a large number of particles. The conditions
at this particular time are such that the energy stored in this excited state
is focused in the back reaction to the field expectation value that acquires
this energy and
whose amplitude will begin to grow.

It is at this point where one recognizes the fundamental necessity of the
``in-in'' formalism in which the equations of motion are real and causal.

\newpage

\section{Conclusions and Discussion}

We have focused on understanding the first stage of a reheating process,
that of dissipation in the dynamics of the field expectation value of a scalar
field via
particle production.
Starting from a scalar field theory with
no apparent dissipative terms in its dynamics, we have shown how the evolution
of expectation value of the scalar field
is affected by the quantum fluctuations and particle production
resulting in dissipative dynamics.

We started our analysis with a perturbative calculation, both in the amplitude
of the expectation value of the field and to one-loop order. Analytically and
numerically we find
that dissipative processes cannot be described perturbatively. A systematic
solution to the equation of motion reveals the presence of resonances and
secular terms resulting in the growth of the corrections to the classical
evolution. The perturbative study of the O(2) linear sigma model reveals
infrared divergences arising from the contribution to the dissipative kernels
from
the Goldstone modes which require non-perturbative resummation.

A Langevin equation was constructed in an amplitude expansion, it exhibits a
generalized fluctuation-dissipation theorem with non-Markovian (memory) kernels
and colored noise correlation functions thus offering a more complex picture of
dissipation.

Motivated by the failure of the perturbative approach, we studied the
non-equilibrium
dynamics in a Hartree approximation both in the symmetric as well as in the
broken
symmetry case. This approximation clearly exhibits the contribution of open
channels and the dissipation associated with particle production, which in this
approximation is a result of parametric amplification of quantum fluctuations.

In the case of unbroken symmetry we find that asymptotically the
expectation value of the scalar field oscillates around the trivial vacuum
with an amplitude that depends on the coupling and {\em initial conditions}.
An extensive numerical study of the renormalized equations of motion was
performed that shows {\em explicitly} the dynamics of particle production
during these oscillations.

Although the Hartree approximation offers a self-consistent non-perturbative
resummation, it is not controlled. Thus we were led to study the large N limit
in an O(N) model, which also allows us to study in a non-perturbative manner
the
dissipative dynamics of Goldstone bosons.
In the case of unbroken symmetry the results are very similar to those obtained
in the Hartree approximation.

The broken symmetry case provides several new and remarkable results.

An extensive numerical study of the equations explicitly shows
copious particle production while the expectation value of the field relaxes
with strongly damped oscillations towards a minimum of the effective action.

It is intuitively
obvious that Goldstone bosons are extremely effective for dissipation since
channels are open for arbitrarily small energy transfer. What is remarkable
is that for ``slow-roll'' initial conditions, that is the initial
expectation value of the scalar field very close to the origin and extremely
small coupling constant (${\cal{O}}(10^{-7})$ or smaller) the {\it final}
value of the expectation value remains very close to the origin, most of the
potential energy has been absorbed in the production of long-wavelength
Goldstone bosons. This is confirmed by an exhaustive numerical study,
including snapshots at different times of the number of particles produced
for different wavelengths
showing a large peak at small wavevectors at long times. The numerical study
clearly shows that
particle production is extremely effective for long-wavelength Goldstone
bosons. Another remarkable result is that the asymptotic value
expectation value of the scalar field
depends on the {\em initial conditions}. These asymptotic
values correspond to minima of the effective action. Thus we reach the
unexpected conclusion that in this approximation the effective action has
a continuum manifold of minima which can be reached from different initial
conditions.

It is pointed out that the basic mechanism for dissipation in this
approximation
is that of Landau damping through the parametric amplification of quantum
fluctuations,
 and production of particles. These  fluctuations react-back in the evolution
 of the expectation value of the scalar field, but out of phase.
This is a collisionless mechanism, similar to that found in the collisionless
Boltzmann-Vlasov equation for plasmas.

Our study also reconciles dissipation in the time evolution for the
coarse grained variable with time reversal invariance, as the evolution is
completely specified by an {\it infinite set} of ordinary second order
differential
equations in time with proper boundary conditions.

Our formalism and techniques are sufficiently powerful to give a great deal of
insight into the particulars of the dissipation process. In particular, we can
see that the damping of the field expectation value ends as the particle
production ends. This
shows that our interpretation of the damping as being due to particle
production
is accurate.

It is useful to compare what we have done here with other work on this issue.
We have already mentioned the work of Calzetta, Hu\cite{calzettahu} and
Paz\cite{paz}. These authors use the closed
time path formalism to arrive at the effective equations of motion for the
expectation value of the field.
Then they solve the {\it perturbative} equations and find dissipative evolution
at short times. In particular Paz finds the kernel that we have found for the
effective equations of
motion of the expectation value in the perturbative and Hartree case. However,
his
perturbative solution is not consistent.

We remedy this situation by studing the non-perturbative Hartree equations,
which
must necesarily be solved numerically as we do.

There has been other, previous, work on the reheating problem, most notably by
Abbott, Farhi and Wise\cite{abbottwise}, Ringwald\cite{ringwald} and
Morikawa and Sasaki\cite{morikawa}. In all of these works, the standard
effective action is used, so that the expectation value is of the ``in-out''
type and hence the equations are non-causal and contain imaginary parts.
In essence, they ``find'' dissipational behavior by adding an imaginary part
to the frequency that appears in the mode equations. We see from our work
(as well as that of Calzetta, Hu and Paz) that this is not necessary;
dissipation can occur even when
the system is evolving unitarily. This comment deserves a definition of what
we call dissipation here: it is the energy transfer from the expectation
value of the $q=0$ mode of the scalar field to the quantum fluctuations
($q \neq 0$) resulting in damped evolution for the $q=0$ mode.

To what extent are we truly treating the reheating problem of inflationary
models? As stated in the introduction, reheating typically entails the decay of
the inflaton into lighter particles during its oscillations.
What we do here is
understand how the quantum fluctuations and the ensuing particle production
influence the dynamics of the evolution  of the expectation value of the
field. Thus technically speaking, this is not the reheating problem.
However, we
{\em are} able to understand where dissipation comes from in a field theory,
and are able to give a quantitative description of the damping process for
the expectation value of the field. We expect that the physics of
dissipation when the
scalar field couple to others will be very similar to the case studied in this
article.

Furthermore, the techniques we develop here are easily adapted to
the case where the inflaton couples to fermions, a case
that we are currently studying\cite{us}. In this connection,
during the writing of this paper, two related pieces of work on the reheating
problem have appeared\cite{lindebranden}. They both look at the effect of
particle production from the oscillations of the inflaton field due to
parametric amplification. What they do {\em not} do is to account for the
back reaction of the produced particles on the evolution of the expectation
value
of the inflaton.
 As we have learned with our study, this back reaction
 will eventually shut off the particle production, so that
these authors may have overestimated the amount or particle production.

We recognize that our non-perturbative treatment neglects the effect of
collisions as mentioned above. Dissipation appears in a manner similar to
Landau damping. In the Hartree approximation there does not seem to be a
natural way to incorporate scattering processes because this is a mean-field
theory.
However the large N expansion allows a consistent treatment of scattering
processes
for which the first contribution (2-2 particle scattering) will appear at
${\cal{O}}(1/N)$. This will be a
necessary next step in order to fully understand the collisional
thermalization which
is the second stage of the reheating process.   This will clearly be a
fascinating and worthy endeavour that we expect to undertake soon.

\section{Acknowledgments}

D. B. would like to thank B. L. Hu, E. Mottola and D. Jasnow
for very illuminating comments and
discussions. He would also like to thank M. Madrid for computational assistance
and
LPTHE for warm hospitality. The authors acknowledge  grants from
the Pittsburgh Supercomputer Center: PHY930049P;
PHY940005P.
D. S. Lee would like to thank Y. Y. Charng for computational help. D. S. Lee
was partially supported by a Mellon Fellowship, D. B. and D. S. Lee were
partially supported by N.S.F. Grant No: PHY-9302534  and N.S.F.
Grant No: INT-9216755 (binational program), they
would like to thank R. Rivers for stimulating conversations.
H. J. de V.  is partially supported by  the CNRS/NSF binational program and
thanks the Dept. of Physics at U. of Pittsburgh for hospitality.
R. H. and A. S. were partially supported by U.S.DOE under contract
DE-FG02-91-ER40682.

\newpage

\section{Appendix I}

In this appendix we give an alternative but equivalent method to derive the
equations of
motion based on the direct time evolution of initial density matrices.
This method gives rise to equations identical in form to those given
in section 2 and 3. However, this method might have greater applicability for
initial
density matrices that are {\em not} of the thermal type, such as might appear
in chaotic inflation models, but is restricted to the gaussian approximation.
For more details the reader is referred to\cite{boyholveg}.

Our starting point is again the Liouville equation for the density matrix
of the system:
\begin{equation}
i \hbar \frac{\partial \rho (t)}{\partial t} = [H_{\rm evol}, \rho(t)],
\label{liouville}
\end{equation}
where $H_{\rm evol}$ is the Hamiltonian of the system that drives the non-
equilibrium evolution of the system.

Next we use the density matrix to extract the order parameter from the field in
the {Schr\"{o}edinger} picture:

\begin{equation}
\phi(t) = \frac{1}{\Omega}\int \ d^3 x\ {\rm Tr} \left[\rho(t)
\Phi(\vec{x})\right],
\end{equation}
with $\Omega$ the spatial volume
(later taken
to infinity), and $\Phi(\vec{x})$ the field in the {Schr\"{o}edinger} picture.
Using
the Liouville equation together with the Hamiltonian:

\begin{equation}
H = \int  d^3 x \left\{\frac{\Pi^2}{2} + \frac{1}{2} (\nabla \Phi)^2 +
V(\Phi) \right\}
\end{equation}
and the standard equal time commutation relations between a field and its
canonically conjugate momentum, we find the equations:

\begin{eqnarray}
\frac{d \phi(t)}{dt} & = & \frac{1}{\Omega}\int d^3x
  \langle \Pi(\vec{x},t) \rangle
 =\frac{1}{\Omega}\int d^3x  Tr \left[ {\rho}(t) \Pi(\vec{x})\right] = \pi(t)
\label{fidot} \\
\frac{d \pi(t)}{dt}  & = & -\frac{1}{\Omega} \int d^3x \langle
 \frac{\delta V(\Phi)}{\delta \Phi(\vec{x})} \rangle . \label{pidot}
\end{eqnarray}

{}From these equations we can find the equation of motion for the order
parameter
$\phi(t)$:
\begin{equation}
\frac{d^2 \phi(t)}{dt^2}+\frac{1}{\Omega}\int d^3x \langle
\frac{\delta V(\Phi)}{\delta \Phi(\vec{x})} \rangle =0 \label{equationofmotion}
\end{equation}

We expand the field operator as $\Phi(\vec{x}) = \phi(t) +
\psi(\vec{x}, t)$, insert this expansion into equation (\ref{equationofmotion})
and keep only the quadratic terms in the fluctuation field $\psi(\vec{x},t)$.
Doing this yields the equation:

\begin{equation}
\frac{d^2 \phi(t)}{dt^2}+V'(\phi(t))+\frac{V'''(\phi(t))}{2 \Omega}\int d^3x
\langle \psi^2(\vec{x},t)\rangle+ \cdots
=0 \label{effequation}
\end{equation}
Here the primes stand for derivatives with respect to $\phi$.

To make sense of the above equation, we need to specify how the expectation
value is to be taken. This entails a specification of the basis in functional
space we will use to write the density matrix in, and, once given this basis,
what the form of the density matrix will be.

In order to generate an expansion in $\hbar$, we choose the density matrix to
be a Gaussian with a covariance of order $1 / \hbar$, in the basis given
by that of the spatial Fourier components of the fluctuation field
$\psi(\vec{x},t)$. Thus write $\psi(\vec{x},t)$ as:

\begin{equation}
\psi(\vec{x},t) = \frac{1}{\sqrt{\Omega}} \sum_{\vec{k}} \psi_{\vec{k}}(t)
e^{-i \vec{k} \cdot\vec{x}}
\label{fourier1}
\end{equation}
If we are choosing the state to be Gaussian, we need to make the Hamiltonian
quadratic, in order that the time evolved state remain Gaussian, thus allowing
for a consistent approximation scheme. To do this we take the Hamiltonian above
and expand it in terms of the fluctuation field out to quadratic order. If we
then insert the expansion in equation (\ref{fourier1})
into the truncated Hamiltonian
obtained in this way, we arrive at:

\begin{eqnarray}
H_{\rm quad} & = & \Omega V(\phi(t))+ \nonumber \\
             &   & \frac{1}{2} \sum_{\vec{k}}
\left\{ -\hbar^2
 \frac{\delta^2}{\delta \psi_{\vec{k}}\delta\psi_{-\vec{k}}}+2
 V'_{\vec{k}}(\phi(t))\psi_{-\vec{k}}+
\omega^2_k(t) \psi_{\vec{k}} \psi_{-\vec{k}} \right\} \label{hamodes} \\
V'_{\vec{k}}(\phi(t))
             & = & V'(\phi(t)) \sqrt{\Omega} \delta_{\vec{k},0} \nonumber \\
\omega^2_k(t)
             & = & \vec{k}^2+V''(\phi(t)). \label{timedepfreq}
\end{eqnarray}

For the density matrix in this basis, we make the following
{\em ansatz}\cite{boyholveg}:

\begin{eqnarray}
\rho[\Phi,\tilde{\Phi},t] & = & \prod_{\vec{k}} {\cal{N}}_k(t)
\exp \left\{- \left[\frac{A_k(t)}{2\hbar}\psi_k(t)\psi_{-k}(t)+
\frac{A^*_k(t)}{2\hbar} \tilde{\psi}_k(t)\tilde{\psi}_{-k}(t)+ \right. \right.
\nonumber \\
                          &   & \left. \left. \frac{B_k(t)}{\hbar}
\psi_k(t)\tilde{\psi}_{-k}(t)\right]
       + \frac{i}{\hbar}\pi_k(t)\left(\psi_{-k}(t)-
\tilde{\psi}_{-k}(t)\right) \right\} \label{densitymatrix} \\
\psi_k(t)
                          & = & \Phi_k - \phi(t) \sqrt{\Omega}
\delta_{\vec{k},0}
\label{psiofkt} \\
\tilde{\psi}_k(t)
                          & = & {\tilde{\Phi}}_k - \phi(t)
\sqrt{\Omega} \delta_{\vec{k},0}
\label{psiprimeofkt}
\end{eqnarray}
where $\phi(t) = \langle \Phi(\vec{x}) \rangle$ and $\pi_k(t)$ is the Fourier
transform of $\langle \Pi(\vec{x}) \rangle$. This form of the density matrix
is dictated by the hermiticity condition $\rho^{\dagger}[\Phi,\tilde{\Phi},t] =
\rho^*[\tilde{\Phi},\Phi,t]$; as a result of this, $B_k(t)$ is real.
The kernel $B_k(t)$ determines the amount of ``mixing'' in the
density matrix, since if $B_k=0$, the density matrix describes a pure
state because it is a wave functional times its complex conjugate.

In the {Schr\"{o}dinger} picture, the Liouville equation (\ref{liouville})
becomes
\begin{eqnarray}
& & i\hbar \frac{\partial \rho[\Phi,\tilde{\Phi},t]}{\partial t} =
\sum_k \left\{
-\hbar^2 \left(
\frac{\delta^2}{\delta \psi_k \delta \psi_{-k}}-
\frac{\delta^2}{\delta \tilde{\psi}_k \delta \tilde{\psi}_{-k}}\right) \right.
 \nonumber \\
& & \left. + {V'}_{-k}(\phi(t))\left(\psi_{k} - \tilde{\psi}_k\right)+
\frac{1}{2}\omega^2_k(t) \left(\psi_{\vec{k}} \psi_{-\vec{k}}
-\tilde{\psi}_k \tilde{\psi}_{-k} \right)
\right\} \rho[\Phi,\tilde{\Phi},t] \label{liouvischroed}
\end{eqnarray}

Since the modes do not mix in this approximation to the Hamiltonian, the
equations for the kernels in the density matrix are obtained by comparing the
powers of $\psi$ on both sides of the above equation. We obtain the following
equations for the coefficients:

\begin{eqnarray}
i\frac{\dot{{\cal{N}}}_k}{{\cal{N}}_k} & = & \frac{1}{2}(A_k-A^*_k)
\label{normeq} \\
i\dot{A}_k                             & = &
\left[ A^2_k-B^2_k -\omega^2_k(t)\right] \label{Aeq} \\
i \dot{B_k}                            & = & B_k (A_k-A^*_k)
\label{Beq} \\
-\dot{\pi}_k                           & = & V'(\phi(t)) \sqrt{\Omega}
\delta_{\vec{k},0} \label{pieq} \\
\dot{\phi}                             & = & \pi
\label{fieq}
\end{eqnarray}
The last two equations are identified with the {\it classical}
 equations of motion for the order parameter (\ref{effequation}).
The equation for $B_k(t)$ reflects the fact that a pure state
$B_k=0$ remains pure under time evolution.

At this point we have not specified any initial conditions; if we were to
specify that the initial state is thermal, we would be led to the following
initial conditions for the kernels $A_k(t),\ B_k(t)$, as well as for the
order parameter and its momentum:

\begin{eqnarray}
A_k(0) & = & A^*_k(0) =
 w_k(0)
\coth \left[\beta_0 \hbar w_k(0)
\right] \label{Ato} \\
B_k(0) & = & - \frac{w_k(0)}
{\sinh\left[\beta_0 \hbar w_k(0)\right]}
\label{Bto} \\
{\cal{N}}_k(0)
       & = &  \left[\frac{w_k(0)}{\pi \hbar}\tanh \left[
\frac{\beta_0 \hbar w_k(0)}{2}\right]\right]^{\frac{1}{2}}
 \label{normo} \\
\phi(0)& = & \phi_0 \; \; ; \; \; \pi(0)=\pi_0
\label{initialfielmom}
\end{eqnarray}
Here $\beta_0$ is the inverse temperature at the initial time (which we have
taken to infinity in this work), while the frequencies $w_k(0)$ are given by:

\begin{equation}
 w^2_k(0) =
\vec{k}^2 +V''(\phi_{cl}(t_o)) \label{adiabfreq}
\end{equation}

Note that up to this point, the formalism we have derived is quite general; it
is only in the specification of the initial conditions that the thermal density
matrix comes in. This is to be contrasted with the closed time path method,
which implicitly assumes a thermal density matrix for the initial state.

Defining a complex function
\begin{eqnarray}
{\cal A}_k (t)   & \equiv & {\cal A}_{kR} (t)+{\cal A}_{kI} (t)\nonumber \\
{\cal A}_{kR}(t) & =      & A_{kR}(t) \tanh \beta_0 \hbar \ w_k(0)
 =  - B_k (t) \sinh \beta_0 \hbar \ w_k(0) \nonumber \\
{\cal A}_{kI}(t) & =      & A_{kI}(t) \nonumber \\
{\cal A}_{kR}(0) & =      & w_k (0) \nonumber \\
{\cal A}_{kI}(0) & =      & 0,
\end{eqnarray}
we find that ${\cal A}_{k}(t)$ satisfies the following Ricatti equation:

\begin{equation}
i\dot{{\cal A}}_k (t) = {\cal{A}}_k (t)^2 - w_k(t)^2
\end{equation}

This equation can be recast in a form closer to that for the mode functions
of section 2 by the following set of transformations:

\begin{eqnarray}
i{\cal{A}}_k (t) & \equiv & \frac{\dot{\chi}_k}{\chi_k} \nonumber \\
\chi_k (t) & \equiv & \frac{1}{\sqrt{w_k(0)}} ({\cal{U}}_{1k} + i
 {\cal{U}}_{2k})
\end{eqnarray}
{}From the thermal initial conditions, and the equation for
${\cal{A}}_k (t)$,
we find that the mode functions ${\cal{U}}_{\alpha  k}$ satisfy the following
equations:

\begin{eqnarray}
\{\frac{d^2}{dt^2} + w_k^2 (t)\}{\cal{U}}_{\alpha k}(t)
                 & = & 0 \nonumber \\
{\cal{U}}_{1k}(0) & = & 1, \; ;  \; \dot{{\cal{U}}}_{1k}(0) = 0 \nonumber \\
{\cal{U}}_{2k}(0) & = & 0, \; ;  \; \dot{{\cal{U}}}_{2k}(0) = w_k(0)
\end{eqnarray}

Finally, using these mode functions to calculate the expectation value of
$\psi(\vec{x},t)^2$ that enters into the equation of motion for the field
expectation value,
we find that equation (\ref{equationofmotion}) becomes:

\begin{equation}
\ddot{\phi}(t) + V'(\phi(t)) + \frac{V'''(\phi(t))}{2} \frac{\hbar}{2}
\int  \frac{d^3k}{(2 \pi)^3} \frac{{\cal{U}}_{1k}(t)^2 +
{\cal{U}}_{2k}(t)^2}{w_k(0)}
\coth \frac{\beta_0 \hbar w_k(0)}{2}=0
\end{equation}
In the limit that the initial temperature goes to zero, these equations are the
same as the one-loop equations derived in section 2.
To obtain the evolutions equations in the Hartree approximation or in the large
N
limit in the ${\cal{O}}(N)$ theory one invokes the Hartree factorization and
proposes a gaussian ansatz as in the one loop case solving for the covariance
following the same steps as above.
We have included this
derivation since it may be applicable to more general situations, and to give
an alternative to the complex path formulation for
the gaussian approximation in non-equilibrium problems .

\newpage

\section{Appendix II}

In this appendix we provide some necessary details that may help the
reader reproduce our numerical results. These concern the numerical integration
for the mode functions $U^+_q (\tau) $ appearing  in various
 models under different approximation schemes. By examing the corresponding
 differential
 equations (for example, eq. (\ref{dimrhartreeequation})) with the
 corresponding boundary conditions a typical integration routine will face
 extremely large derivatives when the values of the wavevectors $q$ reach the
 cutoff.
  We have used improved fourth order Runge-Kutta algorithms for the numerical
  integration, but any algorithm will face the same problems for large enough
cutoff.
  We recognize however, that for large $q$
  (typically of the order of the dimensionless cutoff
 momentum $\approx 100$) for which $q^2 >> 1+\eta^2(\tau)+g \Sigma (\tau)$
   the solutions to these equations are like plane waves with  fast
 varying phases in time.
 It is precisely this fast variation of the phase that  leads to potential
 numerical problems.
 We introduce a WKB-like ansatz
 to take out the fast varying phases
from $U^+_q (\tau)$ by defining  new mode functions which are smoothly
varying in  time for large $q$.
We  consider specifically the $O(N)$ model in the large $N$ limit  as an
example. It is  straightforward to apply this analysis to the other cases
in this paper.

First of all, we define the new mode functions ${\cal{U}}^+_q (\tau)$ as
follows:
\begin{equation}
  {\cal{U}}^+_q (\tau) =e^{i \int_0^{\tau} d{\tau'} \tilde{\omega}_q
({\tau'})} U^+_q (\tau) ~~;~~ \tilde{\omega}_q =\sqrt{q^2+1+\eta^2(\tau)  }
 \label{newmodefunction}
\end{equation}
The initial conditions for ${\cal{U}}^+_q (\tau) $ defined above can be
read off from (\ref{dimbouncon}) for the symmetric phase:
\begin{equation}
   {\cal{U}}^+_q (0) =1 ~~;~~ \dot{\cal{U}}^+_q (0) =0
\label{innewmodefunction}
\end{equation}
By substituting these  ${\cal{U}}^+_q (\tau)$ into the  equations
expressed previously in (\ref{dimrhartreeequation})  we find that the
slowly varying functions obey:

\begin{equation}
  \ddot{\cal{U}}^+_q - 2 i \tilde{\omega}_q (\tau) \dot{\cal{U}}^+_q -i
\dot{\tilde{\omega}}_q (\tau) {\cal{U}}^+_q +g \Sigma (\tau){\cal{U}}^+_q
  =0
\end{equation}

 In can be seen that in the large $q$
 limit, ($q^2 >> 1+\eta^2(\tau)+g \Sigma (\tau)$ ),
\begin{equation}
    {\cal{U}}^+_q (\tau) \rightarrow 1 ~~;~~ \dot{\cal{U}}^+_q (\tau)
\rightarrow 0 ~~;~~ for ~ \tau >0
\end{equation}
which are smooth functions at all times for large $q$.

For the broken symmetry case in $O(N)$ model, the mode
functions ${\cal{U}}^+_q (\tau)$ are solutions of the differential equation

\begin{equation}
  \ddot{\cal{U}}^+_q - 2 i \tilde{\omega}_q (\tau) \dot{\cal{U}}^+_q -i
\dot{\tilde{\omega}}_q (\tau) {\cal{U}}^+_q + \left[ -2+g\Sigma (\tau)
\right]{\cal{U}}^+_q =0
\end{equation}
with the same initial conditions as in (\ref{innewmodefunction}).

\newpage


\newpage

{\bf Figure Captions}

Fig.1: Diagrams contributing to the equation of motion up to
one-loop order, and in the amplitude expansion up to ${\cal{O}}(\phi^3)$.

Fig.2: First order quantum correction for discrete symmetry case $\eta_1(\tau)$
for $\eta_1(0)=0 \; ; \; \dot{\eta}_1(0)=0
\; ; \; \eta_{cl}(0)=1 \; ; \; \dot{\eta}_{cl}(0)=0$. The cutoff is
$\Lambda/m_R =100$.

Fig.3: First order quantum correction for O(2) model $\eta_1(\tau)$ for
$\eta_1(0)=0 \; ; \; \dot{\eta}_1(0)=0
\; ; \; \eta_{cl}(0)=0.6 \; ; \; \dot{\eta}_{cl}(0)=0$. The cutoff is
$\Lambda/\mu_R =100$.

Fig. 4: One loop diagrams contributing to the effective action up to
${\cal{O}}((\phi^{\pm})^4)$. The dashed external legs correspond to the zero
mode.

Fig. 5.a: $\eta(\tau)$ vs $\tau$  in the Hartree approximation, unbroken
symmetry
case. $g=0.1 \; ; \; \eta(0)=1 \; ; \; \Lambda / M_R=100$.

Fig. 5.b: $\Sigma(\tau)$ vs $\tau$ for the same case as in fig. (5.a).

Fig. 5.c: $N(\tau)$ vs $\tau$ for the same case as in fig. (5.a).

Fig. 6.a: $\eta(\tau)$ vs $\tau$  in the Hartree approximation, unbroken
symmetry
case. $g=0.1 \; ; \; \eta(0)=4 \; ; \; \Lambda / M_R=100$.

Fig. 6.b: $\Sigma(\tau)$ vs $\tau$ for the same case as in fig. (6.a).

Fig. 6.c: $N(\tau)$ vs $\tau$ for the same case as in fig. (6.a).

Fig. 7.a: $\eta(\tau)$ vs $\tau$  in the Hartree approximation, unbroken
symmetry
case. $g=0.1 \; ; \; \eta(0)=5 \; ; \; \Lambda / M_R=100$.

Fig. 7.b: $\Sigma(\tau)$ vs $\tau$ for the same case as in fig. (7.a).

Fig. 7.c: $N(\tau)$ vs $\tau$ for the same case as in fig. (7.a).

Fig. 8.a: $\eta(\tau)$ vs $\tau$  in the Hartree approximation, unbroken
symmetry
case. $g=0.05 \; ; \; \eta(0)=5 \; ; \; \Lambda / M_R=100$.

Fig. 8.b: $\Sigma(\tau)$ vs $\tau$ for the same case as in fig. (8.a).

Fig. 8.c: $N(\tau)$ vs $\tau$ for the same case as in fig. (8.a).

Fig. 9.a: $\eta(\tau)$ vs $\tau$  in the Hartree approximation, broken symmetry
case. $g=10^{-5} \; ; \; \eta(0)=10^{-5} \; ; \; \Lambda / \mu_R=100$.

Fig. 9.b: $\Sigma(\tau)$ vs $\tau$ for the same case as in fig. (9.a).

Fig. 9.c: $N(\tau)$ vs $\tau$ for the same case as in fig. (9.a).

Fig. 10.a: $\eta(\tau)$ vs $\tau$  in the large N approximation in the O(N)
model, unbroken symmetry
case. $g=0.1 \; ; \; \eta(0)=1 \; ; \; \Lambda / M_R=100$.

Fig. 10.b: $\Sigma(\tau)$ vs $\tau$ for the same case as in fig. (10.a).

Fig. 10.c: $N(\tau)$ vs $\tau$ for the same case as in fig. (10.a).

Fig. 11.a: $\eta(\tau)$ vs $\tau$  in the large N approximation in the O(N)
model, unbroken symmetry
case. $g=0.1 \; ; \; \eta(0)=2 \; ; \; \Lambda / M_R=100$.

Fig. 11.b: $\Sigma(\tau)$ vs $\tau$ for the same case as in fig. (11.a).

Fig. 11.c: $N(\tau)$ vs $\tau$ for the same case as in fig. (11.a).

Fig. 12.a: $\eta(\tau)$ vs $\tau$  in the large N approximation in the O(N)
model, unbroken symmetry
case. $g=0.3 \; ; \; \eta(0)=1 \; ; \; \Lambda / M_R=100$.

Fig. 12.b: $\Sigma(\tau)$ vs $\tau$ for the same case as in fig. (12.a).

Fig. 12.c: $N(\tau)$ vs $\tau$ for the same case as in fig. (12.a).

Fig. 13.a: $\eta(\tau)$ vs $\tau$  in the large N approximation in the O(N)
model, broken symmetry
case. $g=0.1 \; ; \; \eta(0)=0.5 \; ; \; \Lambda / \mu_R=100$.

Fig. 13.b: $\Sigma(\tau)$ vs $\tau$ for the same case as in fig. (13.a).

Fig. 13.c: $N(\tau)$ vs $\tau$ for the same case as in fig. (13.a).

Fig. 13.d: Number of particles in (dimensionless) wavevector $q$, $N_q(\tau)$
at
$\tau=200$.

Fig. 13.e: Number of particles in (dimensionless) wavevector $q$, $N_q(\tau)$
at
$\tau=13$.

Fig. 13.f: Number of particles in (dimensionless) wavevector $q$, $N_q(\tau)$
at
$\tau=25$.

Fig. 13.g: Number of particles in (dimensionless) wavevector $q$, $N_q(\tau)$
at
$\tau=50$.

Fig. 14.a: $\eta(\tau)$ vs $\tau$  in the large N approximation in the O(N)
model, broken symmetry
case. $g=10^{-7} \; ; \; \eta(0)=10^{-5} \; ; \; \Lambda / \mu_R=100$.

Fig. 14.b: $\Sigma(\tau)$ vs $\tau$ for the same case as in fig. (14.a).

Fig. 14.c: $N(\tau)$ vs $\tau$ for the same case as in fig. (14.a).

Fig. 15.a: $\eta(\tau)$ vs $\tau$  in the large N approximation in the O(N)
model, broken symmetry
case. $g=10^{-12} \; ; \; \eta(0)=10^{-5} \; ; \; \Lambda / \mu_R=100$.

Fig. 15.b: $\Sigma(\tau)$ vs $\tau$ for the same case as in fig. (15.a).

Fig. 15.c: $N(\tau)$ vs $\tau$ for the same case as in fig. (15.a).


\begin{thebibliography}{99}
\bibitem{note} This assumption requires an external force (a Lagrange
multiplier) on the right hand side of (\ref{lineareqofmotion}) for $t<0$
to satisfy the initial conditions. This Lagrange multiplier vanishes for $t>0$,
thus
(\ref{lineareqofmotion}) holds for $t>0$. This assumption corresponds to
``releasing'' the field from the initial value $\delta_i$ with zero initial
velocity and following the time evolution thereafter.
\end{thebibliography}
\end{document}